\documentclass[twocolumn,floatfix,groupedaddress,superscriptaddress,aps,prb,showpacs]{revtex4-1}

\usepackage{graphicx}
\usepackage{amsmath}
\usepackage{amssymb}
\usepackage{amsfonts}
\usepackage{dsfont}
\usepackage{dcolumn}
\usepackage{bbm}
\usepackage{float}
\usepackage{subfigure}

\usepackage{mathtools}

\usepackage{braket}
\usepackage{placeins}
\DeclareMathOperator{\Tr}{\text{Tr}}

\begin{document}

\title{Persisting correlations of a central spin coupled to large spin baths}

\author{Urban Seifert}
\email{urban.seifert@tu-dortmund.de}
\affiliation{Lehrstuhl f\"{u}r Theoretische Physik I, 
Technische Universit\"{a}t Dortmund,
 Otto-Hahn Stra\ss{}e 4, 44221 Dortmund, Germany}

\author{Philip Bleicker}
\email{philip.bleicker@tu-dortmund.de}
\affiliation{Lehrstuhl f\"{u}r Theoretische Physik I, 
Technische Universit\"{a}t Dortmund,
 Otto-Hahn Stra\ss{}e 4, 44221 Dortmund, Germany}

\author{Philipp Schering}
\email{philipp.schering@tu-dortmund.de}
\affiliation{Lehrstuhl f\"{u}r Theoretische Physik I, 
Technische Universit\"{a}t Dortmund,
 Otto-Hahn Stra\ss{}e 4, 44221 Dortmund, Germany}

\author{Alexandre Faribault}
\email{alexandre.faribault@univ-lorraine.fr}
\affiliation{Groupe de Physique Statistique, Institut Jean Lamour (CNRS UMR 7198), Universit\'{e} de Lorraine Nancy, B.P. 70239, F–54506 Vandoeuvre-l\`{e}s-Nancy Cedex, France.}

\author{G\"otz S.\ Uhrig}
\email{goetz.uhrig@tu-dortmund.de}
\affiliation{Lehrstuhl f\"{u}r Theoretische Physik I, 
Technische Universit\"{a}t Dortmund,
 Otto-Hahn Stra\ss{}e 4, 44221 Dortmund, Germany}

\date{\textrm{\today}}

\begin{abstract} 
The decohering environment of a quantum bit is often described by 
the coupling to a large bath of spins. The quantum bit itself can be seen
as a spin $S=1/2$ which is commonly called the central spin. 
The resulting central spin model 
describes an important mechanism of decoherence. We provide mathematically
rigorous bounds for a persisting magnetization of the central spin in this model
with and without magnetic field.
In particular, we show that there is a well defined limit of infinite
number of bath spins. Only if the fraction of very  weakly coupled bath
spins tends to 100\% does no magnetization persist. 
\end{abstract}


\maketitle

\section{Introduction}

In the very active field of coherent quantum control understanding the mechanisms
of decoherence constitutes a major goal. A two-level system or quantum bit is the 
most elementary entity whose coherence is studied. This small quantum system with two-dimensional
Hilbert space can be described as spin $S=1/2$. The environment causing decoherence
may be built from various degrees of freedom. In this study, we focus on
a bath of spins. This commonly considered model bears the name
central spin model (CSM) or Gaudin model in honor of Gaudin who
introduced it in the 1970s as one of the rare cases of integrable quantum
many-body models \cite{gaudi76,gaudi83}.
Aside from this attractive theoretical aspect of the CSM, it indeed describes
a multitude of relevant experimental setups. 
An important example is the electronic spin in a quantum dot
where the spin bath is formed by the nuclear spins of the semiconductor
substrate, for instance, GaAs \cite{schli03,hanso07,urbas13}.
But, also the effective two-level description of the energy levels in a nitrogen vacancy
center in diamond coupled to surrounding $^{13}$C nuclear spins \cite{jelez06,alvar10c}
can be based on the CSM.

For its relevance in describing experimental data the CSM has been also the
subject of a multitude of theoretical investigations of which we can hardly provide
an exhaustive list. Persisting spin polarizations occur in the classical
version of the CSM \cite{merku02,erlin04,alhas06,chen07,stane14b} or in approaches based
on the systematically controlled approximations of master equations 
\cite{khaet02,khaet03,coish04,fisch07,ferra08,coish10,barne11b,barne12}.
Many tools have been developed in the last years
comprising coupled cluster approaches \cite{witze05,yang08a} and 
equations of motion \cite{deng06,deng08}
as well as diagrammatic approaches \cite{cywin09a,cywin09b}.
Heavy numerical approaches are able to simulate baths of about 20 spins $S=1/2$
for long times \cite{talez84,dobro03a,dobro03b,hackm14a} by Chebyshev expansion or up to 1000 spins
for limited times \cite{stane13a}. The existing analytically exact solution
via Bethe ansatz can also be used \cite{bortz07b,bortz10b}, 
but its complexity rises very quickly if
experimentally relevant quantities shall be studied so that only stochastic
evaluations are feasible for bath sizes of up to 48 spins \cite{farib13a,farib13b}.

The caveat of all numerical approaches to the dynamics in the CSM and all approximate analytical
approaches is that they cannot make  statements, which are {\it a priori} reliable, 
about very large spin baths \emph{and} very long times. Only {\it a posteriori} one
may verify whether the results are reasonable or not. 
In principle, analytic results such as general master equations 
are not hampered by constraints in time or system size. But, to our knowledge no
such approach can be evaluated exactly. Generically, an expansion or approximation
related to a small parameter is involved such as the ratio $J_k/b$ of the exchange couplings
$J_k$ over the magnetic field $b$ applied externally to the central spin 
or over some internal polarization \cite{khaet02,khaet03,coish04,coish10,barne12}.
Hence, for small or even vanishing magnetic fields no systematically controlled statements
are possible. Yet, this region is experimentally relevant in spin noise
measurements \cite{li12,kuhlm13,dahba14}.

Thus, it is very useful
to dispose of rigorous results, either to directly interpret experimental data or
to gauge the accuracy of the approximate methods. In spite of the long standing
history of the CSM it was only recently noticed that the generalized Mazur inequality
rigorously shows that persisting correlations are a generic feature of the CSM
if the distribution of couplings is normalizable \cite{uhrig14a}.
Interestingly, however, the physically relevant case of a central electronic spin
with hyperfine couplings to a bath of nuclear spins cannot be normalized because
an infinite number of bath spins couples to the central spin, though most of them only very 
weakly \cite{merku02,schli03,erlin04,chen07}. 
In addition, the temporal fluctuations in the CSM around the 
long-time limit and their evolution has been subject of recent rigorous estimates
in Ref.\ \onlinecite{hette15} 
where the issue of persisting correlations has not been treated.

The goal of this study is three-fold. First, we present rigorous bounds for
very large baths and show that they extrapolate reliably to infinite
bath sizes if this limit can be based on a normalizable distribution of hyperfine couplings. Second, we improve the bounds
obtained previously \cite{uhrig14a}. In zero magnetic field,
the bounds are not yet tight. Third, we extend the rigorous approach
to finite magnetic fields applied to the central spin. Thereby, we enlarge
the applicability of the rigorous approach. By comparison to numerical data, we
illustrate that the rigorous bounds are tight in finite magnetic fields.

The paper is set up as follows. Next, in Sect.\ \ref{sec:modmet}, we introduce the model 
and the employed method in some detail. In Sect.\ \ref{sec:infinite}, we study the
limit of infinite spin baths. In Sect.\ \ref{sec:improved}, we show how the previous
bounds can be improved. In passing, we establish a useful identity to compute 
static spin correlations for infinite spin bath based on Gaussian integrals.
Finite magnetic fields are the focus of Sect.\ \ref{sec:finmag} and the paper
is concluded in Sect.\ \ref{sec:conclusio}.
Technical aspects and the used expectation values are given in the Appendixes.

\section{Model and Method} 
\label{sec:modmet}

\subsection{Model}

\begin{figure}[htb]
\begin{center}
	\includegraphics[width=0.7\columnwidth,clip]{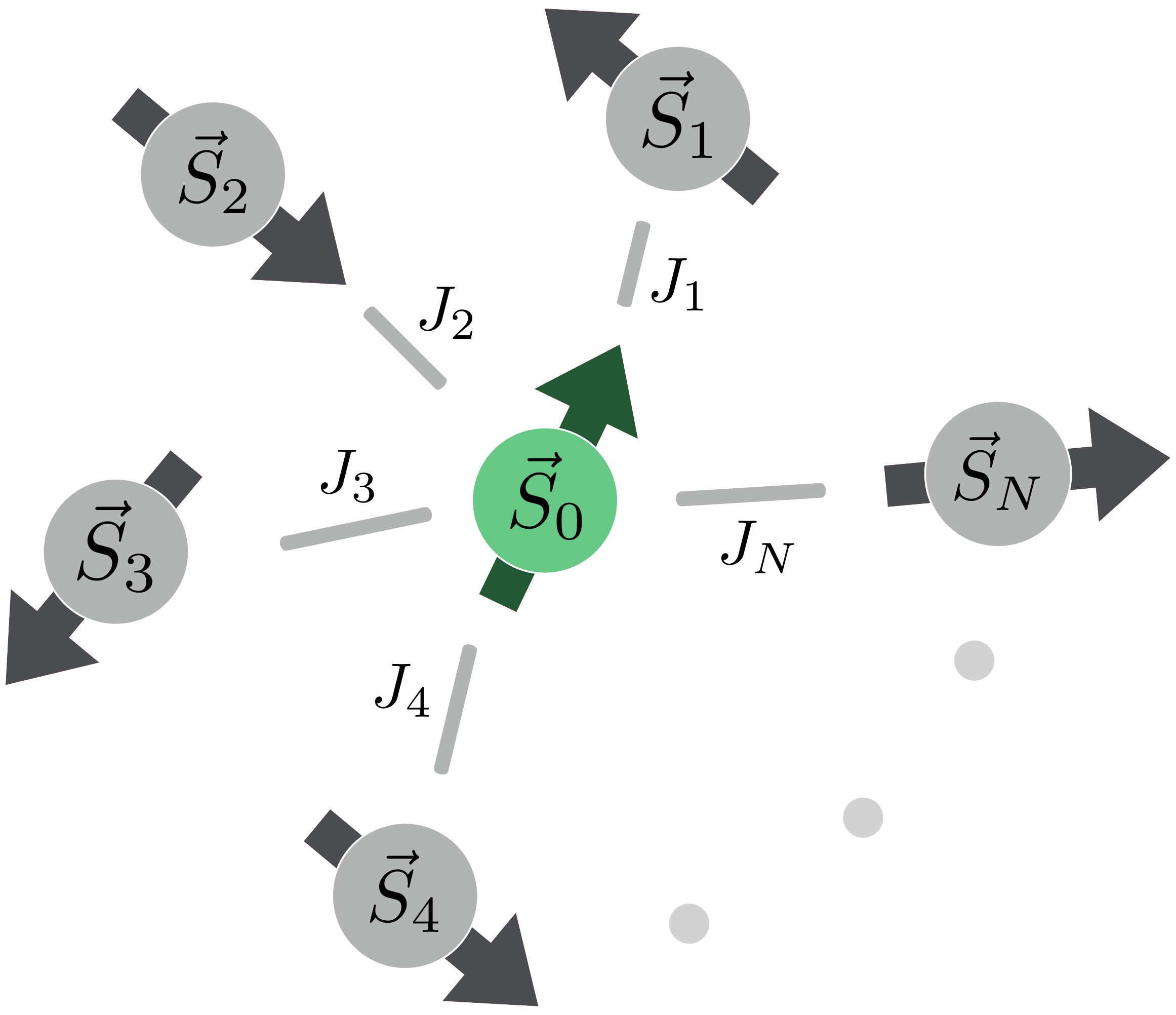}
\end{center}
\caption{(Color online) Scheme of the central spin model (CSM) 
with couplings between the central spin 
$\mathbf{S}_0$ and the surrounding bath spins $\mathbf{S}_k$.
\label{fig:csm}
}
\end{figure}

The CSM is depicted in Fig.\ \ref{fig:csm}.
A central spin interacts with a bath of $N$ surrounding spins.
One may think of the central spin to be an electronic spin, the 
spins of the bath to be nuclear spins, and their coupling to be the
relativistic hyperfine coupling. 
We focus here on the isotropic case  with the Hamiltonian
\begin{equation}
	\label{eq:csm}
	H_0:=\sum_{k=1}^N J_k \mathbf{S}_0 \cdot \mathbf{S}_k,
\end{equation}
where $J_k$ denotes the respective coupling constant of the $k$th bath spin. 
For simplicity, we consider here only $S=1/2$ bath spins. But this restriction
can be relaxed. Moreover, we assume that the $J_k$ are pairwise different
to facilitate the mathematical treatment below.

Applying an external magnetic field with the field strength $h$ to the central spin
leads to
\begin{equation}
	\label{eq:h0h}
	H_0(h):=H_0 - h S_0^z.
\end{equation}
In \eqref{eq:h0h}, we do not include the interaction between the external magnetic field and the bath spins because the magnetic moment of nuclei is typically three orders of magnitude
smaller than the electronic one. But such a term could be considered in an
extended study if needed.

The CSM belongs to the class of integrable Gaudin models \cite{gaudi83}, 
having $N+1$ constants of motion
\begin{equation}
\label{eq:H_l}
  H_l = \sum_{k=0,\neq l}^N \frac{1}{\epsilon_l - \epsilon_k} \mathbf{S}_l \cdot \mathbf{S}_k,
\end{equation}
with $\epsilon_k := - 1/J_k$ and $\epsilon_0 := 0$. Similarly, we also obtain 
$[{H_l(h)},{H_p(h)}]=0$ for $H_l(h):=H_l-hS_l^z$ due to
\begin{eqnarray}
\nonumber
	[{H_l(h)},{H_p(h)}]  &= & - [{H_p},{H_l} ]+
	h [{S_l^z+S_p^z},{\frac{1}{\epsilon_l-\epsilon_p}\mathbf{S}_l \cdot \mathbf{S}_p}]
	\\
	&& -h^2 [{S_p^z},{S_l^z}]
	\label{eq:hlh}
\end{eqnarray}
and the invariance of inner products of vector operators under rotations.

\begin{figure}[htb]
\begin{center}
	\includegraphics[width=1.0\columnwidth,clip]{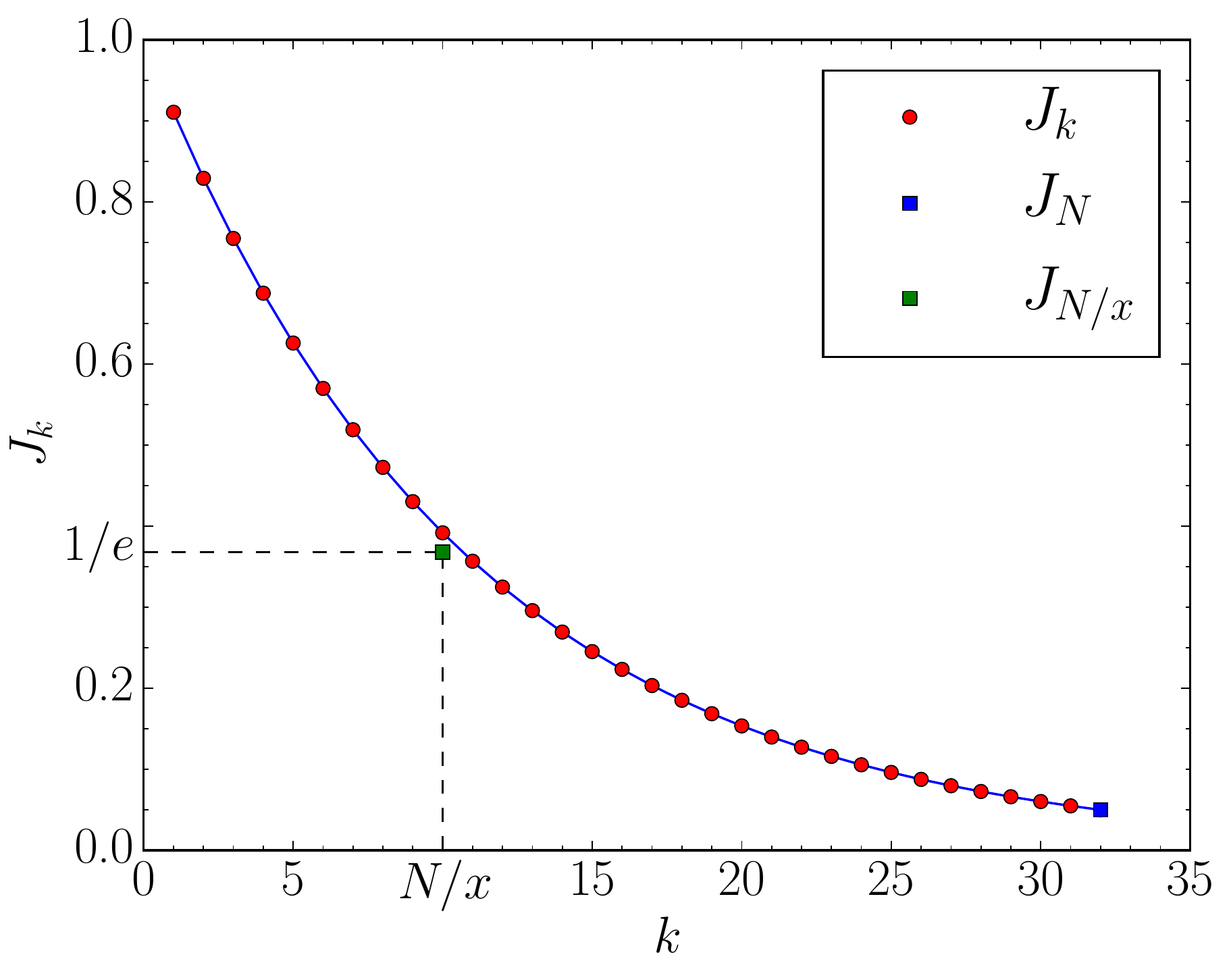}
\end{center}
\caption{(Color online) Example of the exponential coupling distribution defined in 
\eqref{eq:couplings_exp} for $N=32$ bath spins and $x=3$.
\label{fig:distribution}
}
\end{figure}

In quantum dots, the couplings behave as $J_k\propto |\psi(\mathbf{r}_k)|^2$ where 
$\psi(\mathbf{r}_k)$ denotes the electronic wave function of the electron or hole carrying
the central spin at the site of the $k$th nuclear spin \cite{merku02,schli03}. 
For concreteness, we consider the
 following physically reasonable set of couplings \cite{farib13a,farib13b} 
throughout our calculations 
\begin{equation} 
\label{eq:couplings_exp}
  J_k = J \exp \left[-k \frac{x}{N} \right],
\end{equation}
where $x:=N/N_0$ indicates the ratio of the total number of bath spins $N$ to the number of bath spins $N_0$ within the localization radius of the wave function, see
 Fig.\ \ref{fig:distribution}. We refrain from fitting details of coupling distributions
because we are interested in generic features.
The parameter $x$ can be interpreted as controlling the ``spread'' of the couplings $J_k$.
The ratio between the largest coupling $J_1$ and the smallest coupling $J_N$ is given by
$J_1/J_N= \exp(x(1-1/N))$, i.e., small values of $x$ correspond to rather homogeneous
distributions while large values of $x$ correspond to wide-spread distributions.

For further calculations below we define the following moments of the couplings $J_k$ 
\begin{equation}
	\label{eq:sigma}
	\Sigma_m := \sum_{k=1}^N J_k^m
\end{equation}
with $J_Q:=\sqrt{\Sigma_2}$ being commonly used as unit of energy. Of course, the $\Sigma_m$
can be easily computed for the couplings in \eqref{eq:couplings_exp}. But we will use
the $\Sigma_m$ generally below because the bounds can be expressed in 
terms of the $\Sigma_m$.

The coupling constants $J_k$ themselves are in the range of $\mu \mathrm{eV}$
\cite{merku02,lee05,petro08} which corresponds to 
 temperatures of the order of  $10\,\mathrm{mK}$ which are considerably lower than experimentally relevant temperatures \cite{urbas13}. Thus, we assume the bath to be initially
completely disordered so that a density matrix proportional to the identity 
$\rho = \mathds{1} / \Tr [\mathds{1}]$ is used as initial state of the bath 
throughout this paper.

\subsection{Method}
\label{sec:method}

In the previous work Ref.\ \onlinecite{uhrig14a}, a general method was presented
to calculate lower bounds for the autocorrelation function
\begin{equation}
\label{eq:A_correlation}
	A(t) := \braket{\hat A^\dagger(t) \hat A(0)} = \Tr \left[\rho \hat A^\dagger(t) 
	\hat A(0)\right],
\end{equation}
where $\hat A$ is the operator of interest in a system given by the Hamiltonian $H$.
The key idea is to project the operator $\hat A$ onto conserved quantities, also
called constants of motion, 
as much as possible because these projections do not evolve in time.

\begin{figure}[htb]
\begin{center}
	\includegraphics[width=1.0\columnwidth,clip]{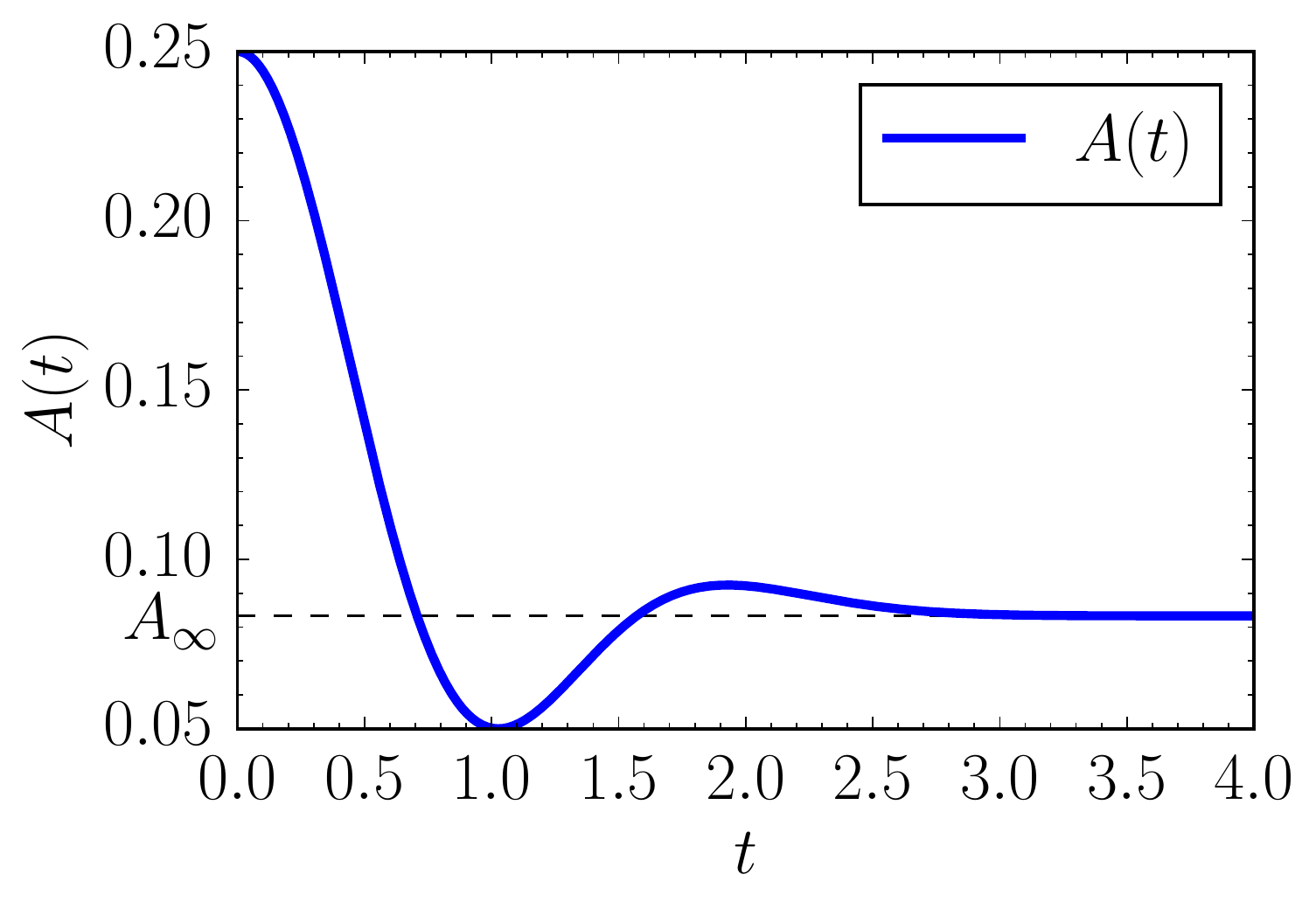}
\end{center}
\caption{(Color online) Example for the autocorrelation function $A(t)$ defined in 
\eqref{eq:A_correlation} with a well-defined limit $A_\infty = \lim_{t \to \infty} A(t)$.
\label{fig:A_t}
}
\end{figure}

If the limit $\lim_{t \to \infty} A(t)$ exists (for an example, see Fig.\  \ref{fig:A_t}), we can 
calculate the lower bound for the long-time limit
\begin{equation}
\label{eq:Ainfty-def}
	A_\infty := \lim_{t \to \infty} \frac{1}{t} \int_0^t A(t^\prime) 
	\ \mathrm{d}t^\prime = \lim_{t \to \infty} A(t).
\end{equation}
Note that there are cases where $A_\infty$ exists as defined in 
\eqref{eq:Ainfty-def} while $\lim_{t \to \infty} A(t)$ does not.
For instance, a purely oscillatory behavior of $A(t)$ around an average value $\overline{A}$
yields $A_\infty=\overline{A}$ while $\lim_{t \to \infty} A(t)$ obviously does not exist.

The lower bound for $A_\infty$ is given by
\begin{equation} 
\label{eq:slow}
  A_\mathrm{low} = \mathbf{a}^\dagger_C \mathbf{N}^{-1} \mathbf{a}_C,
\end{equation}
where the vector $\mathbf{a}_C$ and the matrix $\mathbf{N}$ are given by their respective
elements $a_{C,i} := (C_i|\hat A(0))$ and $N_{im} := (C_i|C_m)$.
The $C_i$ are the conserved quantities, i.e., $[C_i,H]=0$ holds.
The used scalar product of two operators $X$ and $Y$ is defined by
\begin{equation}
	(X|Y) := \braket{X^\dagger Y} = \Tr \left[\rho X^\dagger Y\right]
\end{equation}
We stress that  contrary to the original formulation of Mazur's inequality
\cite{mazur69,suzuk71}, the conserved quantities $C_i$ do not need to be orthonormalized.
Only a matrix inversion of $\mathbf{N}$ is required which can be performed with any computer algebra program.

In the context of the CSM, we discuss the autocorrelation function of the $z$-component of the central spin, i.e., $\hat A = S_0^z$
\begin{equation} 
\label{eq:autocorrelation_function}
	S(t) := \braket{S_0^{z}(t) S_0^z(0)}.
\end{equation}
The expression \eqref{eq:slow} 
with $S_\mathrm{low}:=A_\mathrm{low}$
describes a \emph{lower bound} for the correlation function $S(t)$ which becomes tight if one included the complete set of conserved operators \cite{uhrig14a}.
In our application to the CSM, we try to maximize this bound to make it as tight as possible.
Thus, we consider as many constants of motions as is possible in practice.
Two  interesting aspects arise: (i) which constants of motion contribute significantly to the bound
$S_\mathrm{low}$ and which do not; (ii) which constants of motion contribute
\emph{independently} and which are linearly dependent or close to this.

To judge how tight our bounds are we compare them to numerical data from
Chebyshev polynomial expansions \cite{talez84,dobro03a,dobro03b,hackm14a} and to data from
stochastic evaluation of the Bethe ansatz formulas \cite{farib13a,farib13b}.
Indeed, the model \eqref{eq:csm} is Bethe Ansatz solvable so that every exact eigenstate can be
fully defined in terms of a small set of Bethe roots whose number is, at most, equal to the 
system size $N$. Finding eigenstates in a numerically exact way
then boils down to finding particular
 solutions to a system of $N$ coupled quadratic equations \cite{farib11}, a task which can be
 rapidly carried out for any single arbitrary target state. This allows one to use a simple Metropolis sampling algorithm in order to approximate the observable-specific spectrum, whose
 numerical Fourier transform gives us back the time-evolved expectation value $\left<O(t)\right>$.

\section{Limit of infinite spin bath} 
\label{sec:infinite}

In this section, we deal with the isotropic CSM without magnetic field
and focus on large and infinite bath sizes. Infinite bath sizes are taken
into account via extrapolation.
We consider the exponential coupling distribution \eqref{eq:couplings_exp}
and the $N + 1$ constants of motion 
\begin{subequations}
\begin{align}
	I^z  &:= \sum_{j=0}^N S_j^z
	\\
	H_l^z &:= I^z H_l = \sum_{j=0}^N S_j^z \sum_{k=0,\ne l}^N \frac{1}{\epsilon_l - \epsilon_k} \mathbf{S}_l \cdot \mathbf{S}_k, \quad l \in \mathbb{N}
\end{align}
\end{subequations}
for $N$ bath spins, $l\in \{1,2,3,\ldots N\}$. Note that the
$N+1$ observables $H_l^z$ are linearly dependent; only $N$ of them
are linearly independent \cite{chen07}.

This large number of constants of motion exists thanks to 
the integrability of the system \cite{gaudi76,gaudi83}.
As we will see below the bounds are good, but not tight.

Thus, we also use a mathematically less rigorous route 
which is based on the separation of times scales for large baths.
Instead of the correlation of the central spin one considers the
correlations of the Overhauser field given by
\begin{equation}
\label{eq:overhaus}
\mathbf{B} := \sum_{k=1}^N J_k \mathbf{S}_k.
\end{equation}
Assuming the central spin precesses very rapidly around the Overhauser field
relative to the motion of the Overhauser field itself one can approximate 
 the long-time average of the central spin \cite{merku02} according to 
\begin{subequations}
\begin{align}
	\braket{ \mathbf{S}_0(t)S_0^z(0)} &\approx \left\langle \frac{\mathbf{B}(t)\left( \mathbf{B}(t) 
	\cdot \mathbf{S}_0(t)\right) S_0^z(0)}{B^2(t)} \right\rangle 
	\\
	&\approx \left\langle \frac{\mathbf{B}(t)\left(\mathbf{B}(0) \cdot \mathbf{S}_0(0)\right) 
	S_0^z(0)}{B^2(0)} \right\rangle.
\end{align}
\end{subequations}
where we used the conservation of the Hamiltonian 
$\mathbf{S}(t) \cdot \mathbf{ B}(t) = \mathbf{S}(0) \cdot \mathbf{B}(0)$ 
in the second step. Furthermore, we assume that the
modulus of the Overhauser field $B^2 := |\mathbf{B}|^2$ is conserved. 
This is a good approximation for large baths \cite{merku02}, but not rigorously correct (see 
Sect.\ \ref{sec:id_relevant}).

Then, we can exploit the isotropy of the system and conclude
\begin{subequations}
\begin{align}
	B^\alpha(t)\left(\mathbf{B}(0) \cdot \mathbf{S}_0(0)\right) 
	&= \delta_{\alpha\beta} B^x(t) B^x(0) S_0^\beta (0) 
	\\
	&= \delta_{\alpha\beta} \left(\mathbf{B}(t) \cdot \mathbf{B}(0) \right) \frac{S_0^\beta (0)}{3} 
	\\
	&= \left(\mathbf{B}(t) \cdot \mathbf{B}(0) \right) \frac{S_0^\alpha (0)}{3} . 
\end{align}
\end{subequations}
Finally, we obtain
\begin{equation}
	\braket{ \mathbf{S}_0(t)S_0^z(0)} = \frac{1}{3} \left\langle \frac{\left(\mathbf{B}(t) \cdot\mathbf{B}(0) \right) \mathbf{S}_0(0) S_0^z(0)}{B^2(0)} \right\rangle.
\end{equation}
Focusing on the $z$ component yields
\begin{subequations}
\begin{align}
	\braket{ S_0^{z}(t)S_0^z(0)} &= 
	\frac{1}{3} \left\langle \frac{\left(\mathbf{B}(t) \cdot\mathbf{B}(0) \right) S_0^z(0) S_0^z(0)}{B^2(0)} \right\rangle 
	\\
	&= \frac{1}{12} \left\langle  \frac{\mathbf{B}(t) \cdot \mathbf{B}(0)}{B^2(0)} \right\rangle 
	\\
	&= \frac{1}{12} \frac{S^{(B)}(t)}{S^{(B)}(0)}.
\end{align}
\end{subequations}
Using the isotropy of the model again we restrict ourselves to the autocorrelation function 
$S^{(B)}(t) := \braket{B^{z}(t) B^z(0)}$ of the $z$-component of the Overhauser field
 and to $S^{(B)}(0) = (B^z|B^z)$. Applying \eqref{eq:slow} to $B^z$ yields the estimate
\begin{equation} 
\label{eq:BB}
	S_\text{low,BB} = \frac{S_\text{low}^{(B)}}{12 S^{(B)}(0)}
\end{equation}
which we call the field-field (BB) bound henceforth to distinguish it from the
the spin-spin bound (SS) $S_\text{low,SS} := S_\text{low}^{(S)}$.

\begin{figure}[htb]
\begin{center}
	\includegraphics[width=1.0\columnwidth,clip]{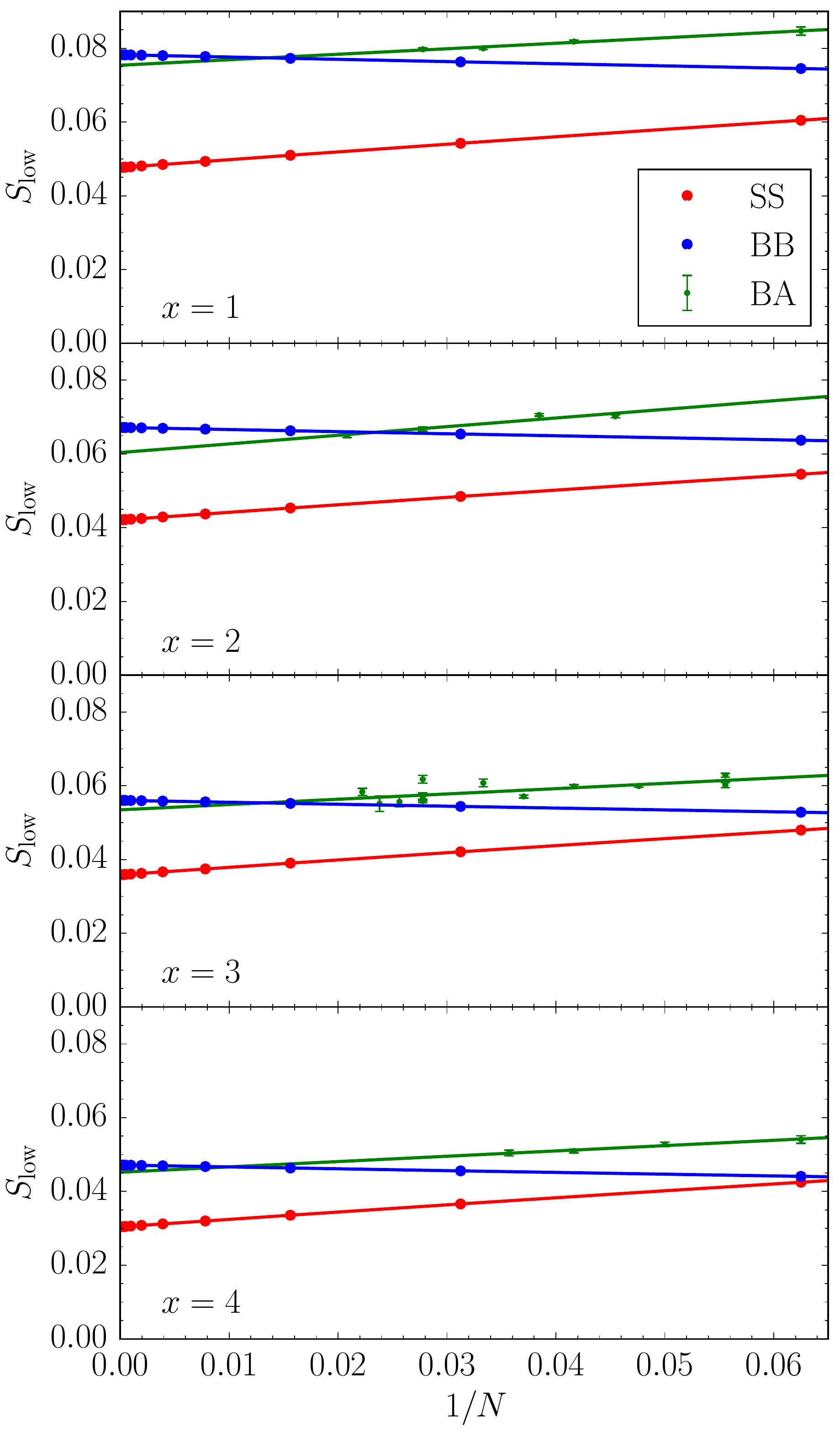}
\end{center}
\caption{(Color online) Lower bounds for $S_\text{low}$ calculated via the spin-spin (SS) and 
via the field-field (BB) autocorrelation for $x \in [1,2,3,4]$ and bath sizes up to 
$N_\text{max} = 4096$. The bounds are compared to results from the stochastic evaluation
of the Bethe ansatz (BA) equations for up to 48 bath spins.
\label{fig:x=1234}
}
\end{figure}

The quantities $S_\text{low}^{(S)}$ and $S_\text{low}^{(B)}$ are the lower bounds calculated for the respective spin-spin and field-field autocorrelation using \eqref{eq:slow}.
The required vector and matrix elements have been calculated previously in 
Ref.\ \onlinecite{uhrig14a}.
We use them here to calculate lower bounds for various bath sizes $N$ 
for arbitrary, but fixed values of $x$ using \eqref{eq:slow}.
The results are shown in Fig.\ \ref{fig:x=1234} for bath sizes up to 
$N_\text{max} = 4096$ for various values of $x$. They are compared to 
data from Bethe ansatz. 

It can be easily seen that results for up to $N_\text{max} = 4096$ bath spins are
 sufficient for a reliable  extrapolation to an infinite spin bath $N \to \infty$ for the spin-spin and for the field-field bounds. This is one of our key results. Our
findings presented in Fig.\ \ref{fig:x=1234} show that
an increasingly denser and denser distribution of couplings implies
a finite thermodynamic limit. We stress, however, that this does not
apply for an increasing spin bath where the ratio $x/N$ is kept constant, i.e.,
where $x$ increases proportionally to $N$.

The observation of an existing thermodynamic limit for given $x$ applies for the 
spin-spin as well as for the field-field bounds. 
But, the bounds for the field-field correlation are remarkably 
tighter in comparison to the exact Bethe ansatz results than
those obtained from the spin-spin correlations. 
Thus, we focus on the field-field bounds below, even though
the spin-spin bounds are rigorous bounds whereas the field-field bounds
involve a physically plausible, but approximate intermediate step.

Technically, we extrapolate the bounds in $1/N$ 
using a cubic polynomial for $x \le 50$ and a quadratic polynomial for $x > 50$.
In addition, we only use data points complying with $N \ge 8x$ to 
guarantee a minimal density of the distributed couplings.
The absolute accuracy of the extrapolations is estimated to be about $10^{-8}$.
This estimate is obtained by (i) comparing extrapolations based on polynomials of 
second, third and fourth degree, (ii) by  varying the number of data points by one or two,
 and (iii) by looking at the standard deviation of the fit parameters. 
The extrapolations based on cubic polynomials are included in 
Fig.\ \ref{fig:x=1234} for the field-field bound and for the spin-spin bound.

For comparison, data obtained by solving the algebraic Bethe ansatz equations using Monte Carlo methods \cite{farib13a,farib13b} and reading off its long-term average 
are included in Fig.\ \ref{fig:x=1234}.
Note that there are multiple data points for a given $x$ and $N$ due to the 
dependence of the data on the starting conditions.
A linear fit of the Bethe ansatz data 
is displayed to obtain the limit $N \to \infty$. 
We decided for a linear fit because of the rather small set of data points and their 
relative scatter. It can be easily seen that the rigorous bounds for the spin-spin correlation do not exhaust the full persisting correlations.
The estimate \eqref{eq:BB} using the field-field correlation works remarkably well 
and might even become exact. But this cannot be decided yet
for lack of accuracy. The accuracy of the Bethe ansatz data is estimated to be about 5$\,\%$.
This is also the range of differences between $S_{\infty,\text{BB}}$ and $S_{\infty,\text{BA}}$.

\begin{figure}[htb]
\begin{center}
	\includegraphics[width=1.0\columnwidth,clip]{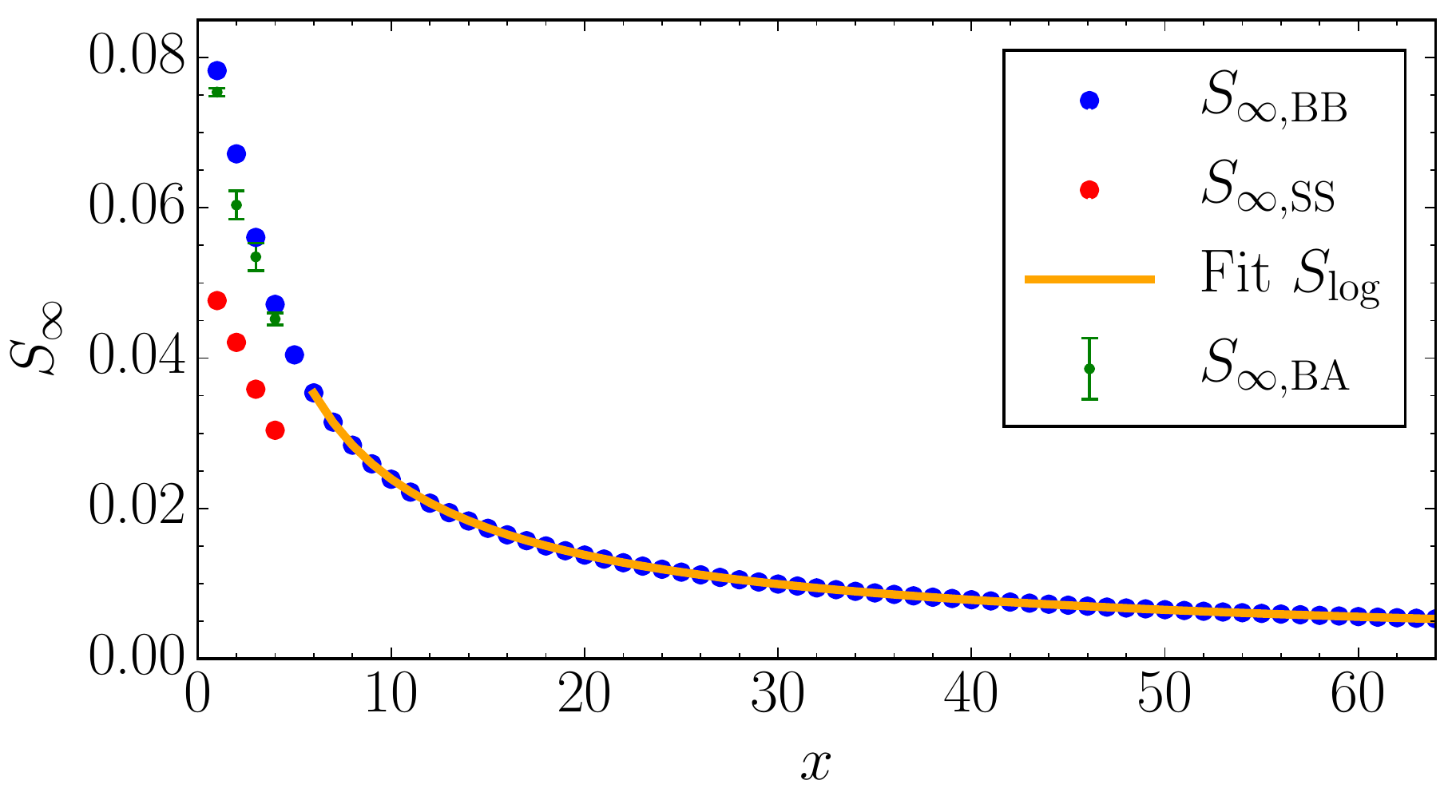}
\end{center}
\caption{(Color online) Data points: extrapolated bounds $S_{\infty,\text{BB}}(x)$ for an infinite spin bath. The solid line is a fit by $S_\text{log}(x)$ from Eq.\ \eqref{eq:s_fit}
in the interval $x \in [6,64]$. We also included the extrapolated values  
$S_{\infty,\text{BA}}(x)$ from the Bethe ansatz data and 
$S_{\infty,\text{SS}}(x)$ from the spin-spin bounds for $x \in [1,2,3,4]$.}
\label{fig:dynamic}
\end{figure}

Figure\ \ref{fig:dynamic} displays the extrapolated bounds relevant for infinite
spin baths as they depend on the spread value $x$. One clearly sees that the
persisting correlation tends to zero for larger and larger spread.
To understand better how $S_\infty(x)$ behaves we consider the simple
estimate from Eq.\ 11 in Ref.\ \onlinecite{uhrig14a} given as
\begin{equation}
S_\mathrm{low} = \frac{1}{4}\frac{\Sigma_1^2}{2\Sigma_1^2+3(N-1)\Sigma_2}
\end{equation}
and insert 
\begin{equation}
\Sigma_m = J^m \frac{N}{m x}(1-\exp(-mx))
\end{equation}
resulting from \eqref{eq:couplings_exp} in the limit $N\to\infty$. This yields
\begin{equation}
S_\mathrm{low} = \frac{1}{6x}\frac{(1-\exp(-x))^2}{1-\exp(-2x)}
\end{equation}
which clearly shows the proportionality $S_\mathrm{low} \propto 1/x$. Thus, we analyze
our more elaborate results in Fig.\ \ref{fig:dynamic} in a very similar way. 
By various fits we find that the power law $\propto 1/x$ fits best, but not
perfectly. Some slowly varying corrections are present and we check them to 
be logarithmic with arbitrary exponent  $S_\mathrm{low} \propto \ln(x)^\alpha/x$.
It turns out that $\alpha=1$ fits very nicely. Thus, we finally test
\begin{equation}
\label{eq:s_fit}
	S_\text{log}(x) = \frac{A \cdot \ln \left(\frac{x}{B}\right)}{x}
\end{equation}
and fit this formula to our data within the intervals $x \in \left[x_\text{start},64\right]$.
The resulting parameters are listed in Table \ref{tab:parameters}. 
Indeed, the two parameters do not change much and appear to converge for 
increasing $x_\text{start}$. Additionally, the fit included in Fig.\ \ref{fig:dynamic}
underlines the impressive agreement so that we conclude that \eqref{eq:s_fit}
describes the asymptotic evolution with $x$ correctly.

In this context, we draw the reader's attention to the heuristic argument
by Chen and co-workers stating that up to time $t$ only those spins
of the bath really contribute to the dynamics which are sufficiently coupled
\cite{chen07}. Therefore, there is an effective bath size $N_\text{eff}(t)$
defined by $t J_{N_\text{eff}}\approx 1$ implying $x(t) \propto \ln(t)$. 
So our finding for the asymptotic dependence \eqref{eq:s_fit}
implies the very long-time behavior $S(t) \propto \ln(\ln(t))/\ln(t)$.
The dominant inverse logarithm has been found previously 
in many studies \cite{khaet02,khaet03,coish04,erlin04,chen07} so that our result
agrees and confirms this point. In addition,
it refines the previous claims on the long-time behavior 
which did not include the nested logarithm in the numerator.

\begin{table}[htbp]
  \centering
      \caption{Parameters of the fit $S_\text{log}(x)$ in \eqref{eq:s_fit} to our data 
			$S_{\infty,\text{BB}}(x)$ within the interval $x \in \left[x_\text{start},64\right]$.}
    \begin{tabular}{c|c|c}
    \hline
    $x_\text{start}$ & $A$   & $B$ \\
    \hline
    6     & 0.05345 $\pm$ 0.00009 & 0.1141 $\pm$ 0.0009 \\
    10    & 0.05426 $\pm$ 0.00007 & 0.1235 $\pm$ 0.0008 \\
    14    & 0.05473 $\pm$ 0.00004 & 0.1294 $\pm$ 0.0005 \\
    18    & 0.05498 $\pm$ 0.00003 & 0.1328 $\pm$ 0.0004 \\
    24    & 0.05519 $\pm$ 0.00002 & 0.1357 $\pm$ 0.0003 \\
    30    & 0.05531 $\pm$ 0.00001 & 0.1374 $\pm$ 0.0002 \\
    \hline
    \end{tabular}%
  \label{tab:parameters}%
\end{table}

As the key result of this section, we highlight the existence of persisting correlations 
in the CSM for the limit $N \to \infty$ with a fixed $x$. The rigorous spin-spin bounds obtained
 by using the extensive number of constants of motion in the CSM do not exhaust the persisting correlations as shown by the comparison to Bethe ansatz data. 
But, the approximate estimate \eqref{eq:BB} using the field-field correlation yields very 
promising results which appear to be rather tight. 
They are used to discuss the extrapolated thermodynamic values $S_{\infty,\text{BB}}(x)$ 
for various $x$. We analyzed the asymptotic values $S_\text{log}(x)$ as shown in 
Fig.\ \ref{fig:dynamic}. 

Our findings suggest that any normalized distribution of couplings $p(J)$ with $\int p(J)dJ=1$
has a well-defined limit $N\to\infty$. It is the infinite number of 
almost uncoupled spins in the exponentially parametrized couplings \eqref{eq:couplings_exp} 
which spoils the persisting correlations.

At present, we cannot decide wether the field-field bounds are still rigorous bounds
since their derivation involves a plausible, but approximate step.
Hence, we try to maximize the lower bounds from the rigorous spin-spin correlation 
in the following section by taking further combinations of conserved operators into account.

\section{Improved bounds for spin-spin correlations} 
\label{sec:improved}

In this section, we aim at improving the rigorous bound for the spin-spin correlations
in the CSM without magnetic field. The goal is to make these bounds as tight as possible.
We first identify relevant conserved quantities and present a method to determine exact analytical expressions for arbitrary matrix elements. Finally, we derive results for the limit of large baths
 $N\to \infty$. 

\subsection{Identifying relevant conserved quantities}
\label{sec:id_relevant}

In order to apply the method presented in Sect.\ \ref{sec:method}, one needs to know vector and matrix elements which are scalar products of two operators $A$ and $B$ being
sums of products of spin operators. Hence, the scalar product takes the form
\begin{equation} 
\label{eq:general_scalarProduct}
  (A | B) =\sum_{i=1}^N \dots \sum_{k=1}^N \Tr \left[\rho S_i^\alpha, \dots, S_k^\beta \right] J_i \dots,
\end{equation}
where we denote spin components $\{x,y,z\}$ with Greek indices for which the sum convention is used. The density matrix $\rho$ is proportional to the identity and normalized. 
For sums over the spins 
\begin{equation} 
\label{eq:spinoperator}
  S_k^\alpha = \mathds{1}_0 \otimes \dots \otimes \sigma_k^\alpha / 2 
	\otimes \dots \otimes \mathds{1}_N  
\end{equation}
Latin indices are used. Depending on the prefactors in the conserved quantity, we may also 
sum over the coupling constants $J_i$.

Since the Pauli matrices are traceless, at least two spin operators, see \eqref{eq:two_contract}
below, need to act on the same spin $k$ in order to generate a non-vanishing contribution in 
\eqref{eq:general_scalarProduct}.
It is also possible to contract more than two operators at one site, 
see \eqref{eq:three_contract},
\begin{subequations}
\begin{align}
  \sum_{k,j=1}^N &\Tr \left[\rho S_k^\alpha S_j^\alpha \right] = \sum_{k=1}^N \Tr 
	\left[\rho S_k^\alpha S_k^\alpha \right] = 3 N / 4 
	\label{eq:two_contract} 
	\\
  \sum_{k,j,j'=1}^N &\Tr \left[\rho S_k^x S_j^y S_{j'}^z \right] = \sum_{k=1}^N \Tr 
	\left[\rho S_k^x S_k^y S_k^z \right] = i N / 8. 
	\label{eq:three_contract}
\end{align}
\end{subequations}

Each sum over the site indices yields a contribution proportional to $N$. Thus, each contraction
yields a factor $N$ and so the most important contributions stem from 
the terms with the maximum number of contractions. This still holds if the scalar product also contains operators such as $\sum_{k}^N J_k\mathbf{S}_k$ so that the sum yields the moments
 $\Sigma_m$.
We assume that the couplings $J_k$ are chosen from a normalized 
distribution $p(J)$ such that all moments exist. Then $\Sigma_m \propto N$ holds.
Therefore, it remains true that the highest order in $N$ results from the maximum 
number of contractions. This is achieved if in each contraction only a minimum number of
operators are contracted, hence, the pairwise contractions \eqref{eq:two_contract} 
yield the most significant  contribution in powers of $N$.

We aim at finding bounds for the autocorrelation function of $S_0^z$.
Let us consider an arbitrary operator $A$ composed of products of $m$ spin operators 
from the bath and each of these spin operators is summed over the whole bath.
Then, the vector element $(S_0^z | A)$ is proportional to 
$N^{\left \lfloor{m/2}\right \rfloor}$ because the maximum number
of contractions can only be  formed from pairs of spin operators in $A$.
If there are an odd number of spin operators, i.e., $m$ is odd, one contraction 
needs to include three operators  yielding a factor $N$. This is why the
largest integer $\left \lfloor{m/2}\right \rfloor$ not larger than $m/2$ defines 
the maximum power in $N$.

The maximum power in $N$ of the norm $(A|A)$ is  $N^m$
because one can form $m$ pairs of spins from the $2m$ spin operators 
occurring in $A^\dagger A$.
If $m$ is even, i.e.,  $m=2k$ with $k \in \mathbb{N}$, we see that
\begin{subequations}
\begin{align}
  S_\mathrm{low} &= \frac{\left(\mathcal{O}(N^{\left \lfloor{m/2}\right \rfloor})\right)^2}
	{\mathcal{O}(N^m)} 
	\\
  &= \frac{\left(\mathcal{O}(N^k) \right)^2}{\mathcal{O}(N^{2k})} = \mathcal{O}(1),
\end{align}
\end{subequations}
This means that such an operator $A$ provides a lower bound which is relevant 
even for infinitely large bath. We deduce from this consideration that such operators,
i.e., operators with even number of summed spin operators, yield finite contributions
to the lower bounds even in the thermodynamic limit.

If, however, $m$ is odd with $m = 2k+1$ ($k\in \mathbb{N}$), we have 
$\left \lfloor{m/2}\right \rfloor = k$ and $(A|A) = \mathcal{O}(N^{2k+1})$.
This results in $S_\mathrm{low} = \left(\mathcal{O}(N^k)\right)^2 / \mathcal{O}(N^{2k+1}) = \mathcal{O}(1/N)$, and hence $S_\mathrm{low} \to 0$ for $N \to \infty$.
We therefore conclude that lower bounds using constants of motion with an \emph{even} number of summed spin operators stay finite for $N \rightarrow \infty$ while constants of motion with an \emph{odd} number of summed spins yield lower bounds which vanish for large $N$.

Furthermore, we stress that a relevant conserved quantity needs to be a vector component in order to have a non-vanishing overlap with $S_0^z$.
This can be achieved by multiplying scalar quantities with a vector component, e.g.,
 $H_0$ with $I^z$ as already done in Ref.\ \onlinecite{uhrig14a}.

Based on these criteria, we identify $I^z H_0^{2m-1}$ with 
$m \in \mathbb{N}$ and $I^z I^2 H_0$ as relevant conserved quantities.
We use the definition
\begin{equation}
\label{eq:total-momentum}
\mathbf{I} := \sum_{k=0}^N \mathbf{S}_k
\end{equation}
for the vector operator of the total momentum.

\citeauthor{merku02} argued in Ref.\ \onlinecite{merku02} that the modulus of the  Overhauser field 
$B^2 = \sum_{k,l=1}^N J_k J_l \mathbf{S}_k \cdot \mathbf{S}_l $ is a conserved quantity as well.
We note, however, that $B^2$ can only be considered conserved in the approximation that the 
long-time average of the central spin is relevant for dynamics of the bath spins.
Rigorously, using the integrability of the CSM with $H_l$ defined in \eqref{eq:H_l} one can write
\begin{subequations}
  \begin{align} 
	\label{eq:comm_hl}
  \sum_{l=1} H_l J_l &= \sum_{l=1} \sum_{j=0,\neq l} 
	\frac{1}{\epsilon_l - \epsilon_j} \frac{-1}{\epsilon_l} \mathbf{S}_l \cdot \mathbf{S}_j 
	\\
    &=  \underbrace{\sum_{l=1} \sum_{j=1,\neq l} \frac{1}{\epsilon_l - \epsilon_j} 
		\frac{-1}{\epsilon_l} \mathbf{S}_l \cdot \mathbf{S}_j}_{=: \zeta} + 
		\underbrace{\sum_{l=1} \frac{-1}{\epsilon_l^2} \mathbf{S}_0 \cdot \mathbf{S}_l}_{=: -\eta}, 
  \end{align}
\end{subequations}
The first term $\zeta$ reduces to 
\begin{equation}
\zeta =  B^2 / 2 - \frac{3}{8} \Sigma_2 \mathds{1}
\end{equation}
 so that $[{H_0},{\zeta}] = [{H_0},{B^2}]/2$. For the second term, we find
\begin{equation}
  [{H_0},{\eta}] = \frac{i}{2} \sum_{k=1} \sum_{l =1, \neq k} \left(J_k J_l^2 - J_l J_k^2 \right) \epsilon_{\alpha \beta \gamma} S_0^\alpha S_k^\beta S_l^\gamma,
\end{equation}
which does not vanish except for the particular case of uniform
couplings. Since the left hand side of \eqref{eq:comm_hl}
is conserved, the last equation implies that $Q := B^2 - 2 \eta$ is conserved
so that one may use $Q$ instead of $B^2$ as constant of motion. 
But, an explicit calculation reveals  that $H_0^2$ and $Q$ are identical up to
a prefactor 
\begin{subequations}
\begin{eqnarray}
\label{eq:Bsquared}
  \left( \mathbf{S}_0 \cdot \mathbf{B} \right) \left( \mathbf{S}_0 \cdot \mathbf{B} \right) 
	&=& 
	\frac{1}{4} B^2 + \frac{i}{2} \mathbf{S}_0 \cdot \left( \mathbf{B} \times \mathbf{B} \right)
	\\
	&=& \frac{1}{4} Q.
	\end{eqnarray}
	\end{subequations}
Therefore, the inclusion of $Q$ does not yield any additional insight compared to powers
of $H_0$.

\subsection{Computer aided analytics for finite spin baths} 

\label{sec:computer_aided}
\subsubsection{Method}

The calculation of the analytical expressions for matrix elements by hand becomes more and more tedious for  increasing number of spin operators. For this reason, we resort to a computer-aided approach which we sketch here. We consider scalar products of the form
\begin{equation} 
\label{eq:scalarpr_allg}
	(A|B)=\Tr[\rho A^\dagger B]
\end{equation}
where $A$ and $B$ contain sums over spin operators of the bath
weighted or not with the couplings $J_k$. In addition, the magnetic field
applied to the central spin may occur. Then we know from the above arguments
for contractions that the general result is given in the form of a sum over
monomials consisting of powers of $N$, the moments $\Sigma_m$, and the magnetic field $h$.
We denote these monomials by $f_i(\mathcal{J},h)$ where $\mathcal{J}$ stands for the
set of coupling $\{J_k\}$ which defines the moments $\Sigma_m$. But we stress that 
the dependence of $(A|B)$ on $\mathcal{J}$ enters via the moments $\Sigma_m$.
Thus, we start from the ansatz
\begin{equation} 
\label{eq:scalar_ansatz}
	(A|B)=\sum_{i=1}^M \alpha_i f_i(\mathcal{J},h).
\end{equation}
The number $M$ and the precise form of possible monomials can be estimated beforehand
by general considerations such as the units and the minimum and the 
maximum number of possible contractions.
We illustrate this point in a concrete example below.
The task we hand over to the computer is to compute the coefficients $\alpha_i$
which generally are fractions. Note that the coefficients enter
linearly in the equations of type \eqref{eq:scalar_ansatz}.
We implement an algorithm to calculate the traces 
needed to determine scalar products for concrete, rather small
baths of up to six spins for $M$ sets of couplings  $\mathcal{J}$.
 Two possible approaches are outlined in Appendix \ref{app:algos}.

In this way, we obtain a system of $M$ linearly independent equations with $M$ variables 
$\alpha_i$ for $M$ different explicit choices of  $\mathcal{J}$ and 
magnetic field strengths $h$.
Some of the couplings may be set to zero which amounts to considering smaller
baths. This procedure yields the set of $M$ linear equations
\begin{equation}
	\label{eq:lgs}
	\sum_{i=1}^M \alpha_i f_i(\mathcal{J}_m,h_m)=\Tr[\rho A^\dagger B]_m,~m=1,...,M .
\end{equation}
The solution of this set of equations \eqref{eq:lgs}  yields the desired 
coefficients $\alpha_i$. The results can be checked for 
additional sets $\mathcal{J}_m, h_m$ with $m>M$.

As an example we consider the non-diagonal matrix element $(I^z | I^z H_0^2)$ for $h=0$.
This element is quadratic in $H_0$ so that its monomials must have
units quadratic in energy. 
Furthermore, there are  two operators $I^z$ which comprise sums over the bath, 
but not over the coupling constants.
If the spins in these two operators are contracted, we obtain an explicit factor of $N$.
Thus, the potentially relevant monomials $f_i$ are $N \Sigma_2$, $\Sigma_2$ and $\Sigma_1^2$.
No other monomial matters.
Considering three sets of couplings $\mathcal{J}_1, \mathcal{J}_2$ and $\mathcal{J}_3$, 
we obtain three equations of the form
\begin{equation}
   \alpha_1 (N \Sigma_2) (\mathcal{J}_i) + \alpha_2 \Sigma_2(\mathcal{J}_i) + 
	\alpha_3 \Sigma_1^2 (\mathcal{J}_i) = (I^z | I^z H_0^2) (\mathcal{J}_i).
\end{equation}
For example, one can consider $\mathcal{J}_1=\{1\}$, $\mathcal{J}_2=\{1,1\}$ and 
$\mathcal{J}_3 = \{1,2\}$, yielding the  results $(I^z | I^z H_0^2)(\mathcal{J}_1) = 2/64$, 
$(I^z | I^z H_0^2)(\mathcal{J}_2) = 14/64$ and $(I^z | I^z H_0^2)(\mathcal{J}_3) = 33/64$.
The resulting coefficients read as $\alpha_1 = 3/64$, $\alpha_2=-3/64$ and $\alpha_3=2/64$.
The appearance of the denominator $2^6$ stems from the simple fact that  $(I^z | I^z H_0^2)$
is made from products of six spin operators of $S=1/2$. The computer-aided calculations
can be implemented such that the numbers in the matrices and vectors are integers
so that no rounding errors occur at all.
The choice of the sets of couplings $\mathcal{J}$ is arbitrary except that the resulting
equations must  be linearly independent.
We emphasize that the applicability of this approach is only limited by the runtime of the
 algorithm used to compute the numerical results for the trace evaluations.
For further algorithmic details, the reader is referred to Appendix \ref{app:algos}.

\subsubsection{Results for zero magnetic field}

\begin{figure}[htb]
  \includegraphics[width=1\columnwidth,clip]{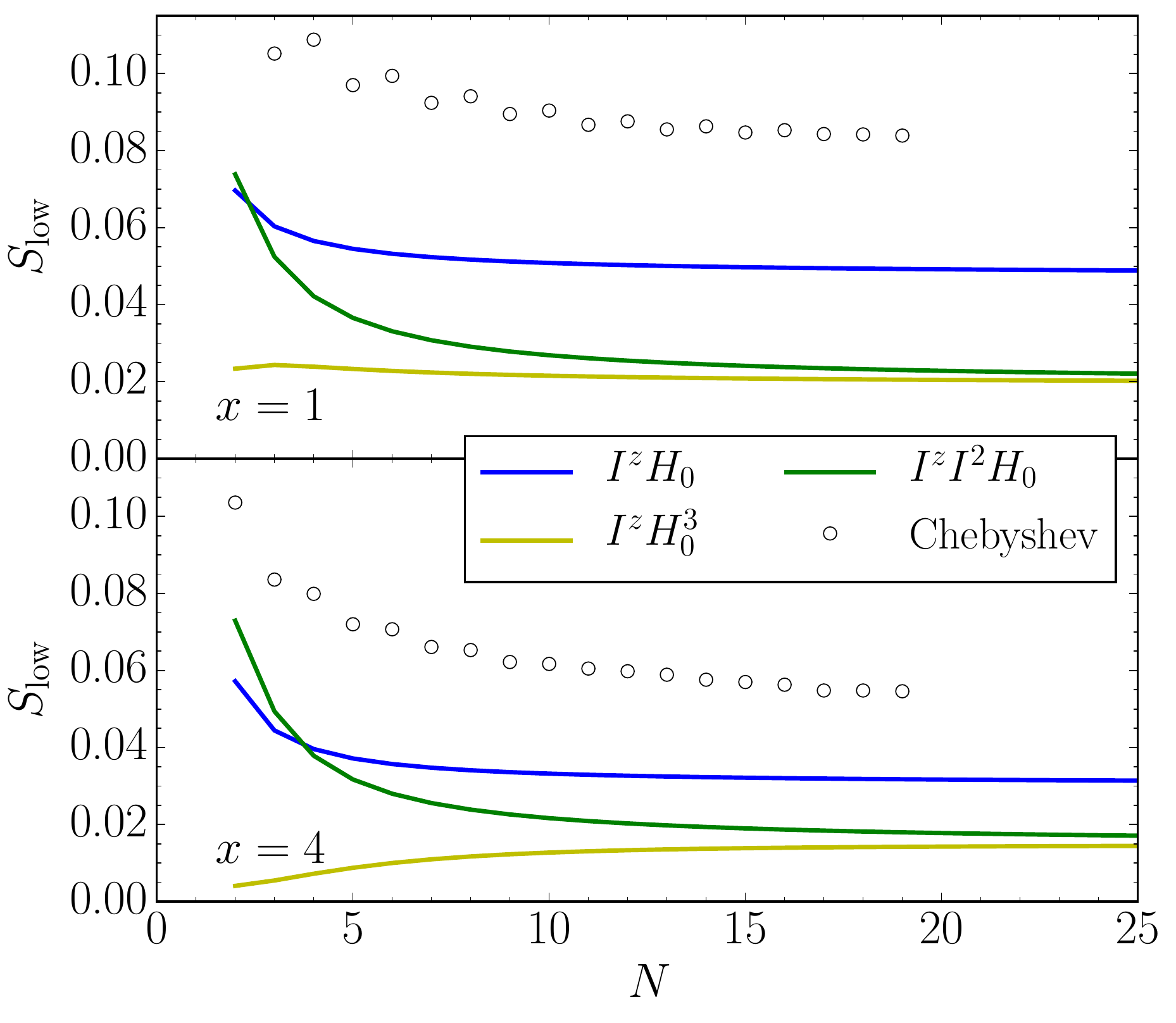}
\caption{(Color online) Comparison of lower bounds using single conserved quantities with an 
even number of summed spins so that $S_\mathrm{low}$  remains finite for
$N\to\infty$. The coupling distribution is given by \eqref{eq:couplings_exp}.
To show how tight the bounds are, Chebyshev expansion data
is included \cite{hackm14a,pb_hackmann_2015}.
Upper panel: $x=1$; lower panel: $x=4$.
\label{fig:compare_singles}}
\end{figure}

\begin{figure}[htb]
  \includegraphics[width=1\columnwidth,clip]{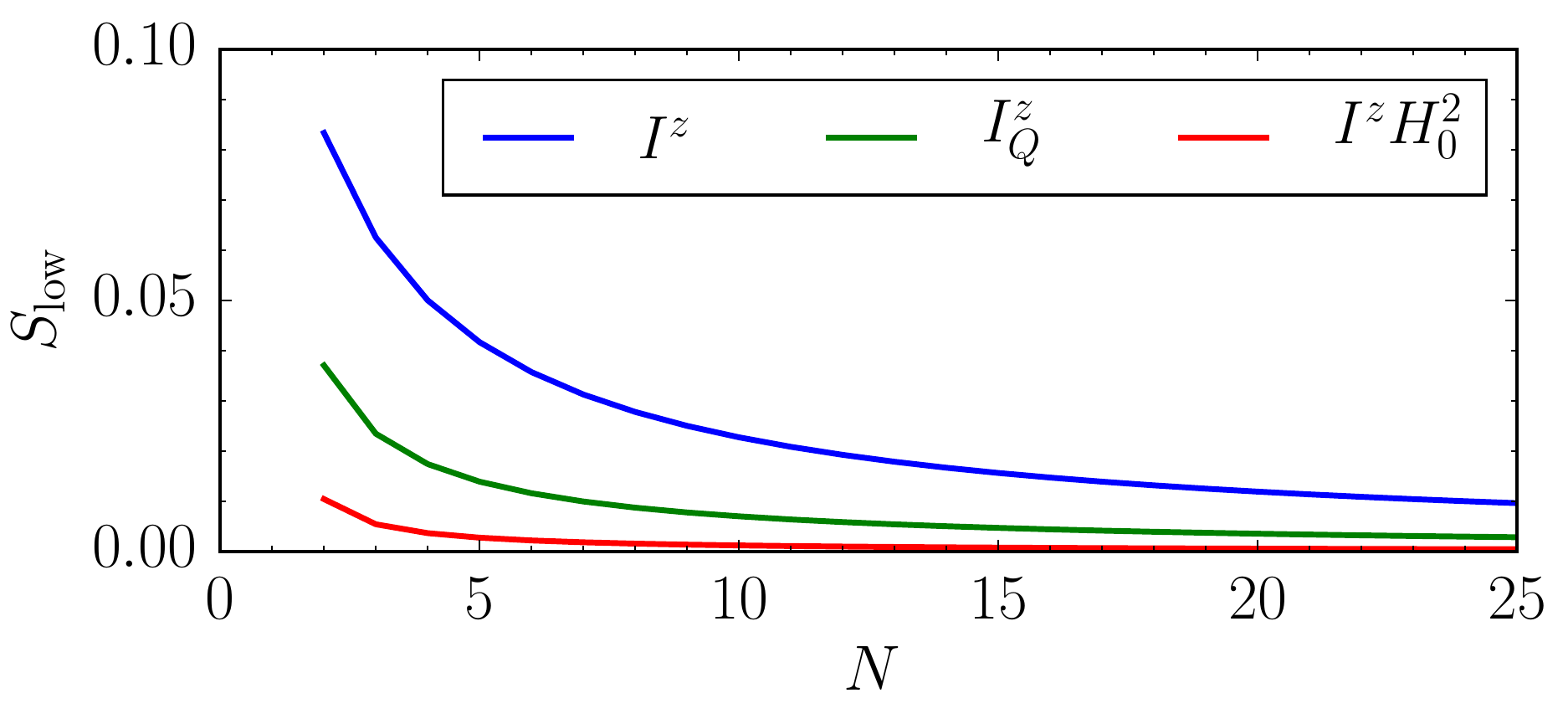}%
 \caption{(Color online) Comparison of lower  bounds resulting from single conserved quantities
 with an odd number of summed spins so that $S_\mathrm{low} \to 0$ holds. The coupling distribution is given by \eqref{eq:couplings_exp} with $x=4$. We point out
that the lower bounds for $I^z$ and $I_Q^z$ do not depend on the  couplings at all.
The significance of the lower bound from $I^z H_0^2$ is rather low.
\label{fig:compare_singles_vanishing}}
\end{figure}

We apply the approach sketched above to calculate the matrix and vector elements for 
$I^z H_0^3$ and $I^z I^2 H_0$ as well as for $I^z H_0^2$ although we do not expect significant improvements from the last operator, see Sec.\ \ref{sec:id_relevant}.
The required input of the matrix elements is listed in Appendix \ref{app:elements_list}.
We evaluate these elements for the exponential distribution of 
couplings \eqref{eq:couplings_exp}.
In Figs.\ \ref{fig:compare_singles} and \ref{fig:compare_singles_vanishing}, the lower bounds 
using only one conserved quantity are shown.
The asymptotic behavior for $N\to\infty$ is clearly discernible.
In perfect agreement with our analytical reasoning in Sec.\ \ref{sec:id_relevant},
conserved quantities with \emph{even} number of summed spins yield a 
finite lower bound, see Fig.\ \ref{fig:compare_singles},
 while quantities with an \emph{odd} number of 
summed spins vanish, see Fig.\ \ref{fig:compare_singles_vanishing}.

\begin{figure}[htb]
  \includegraphics[width=1\columnwidth,clip]{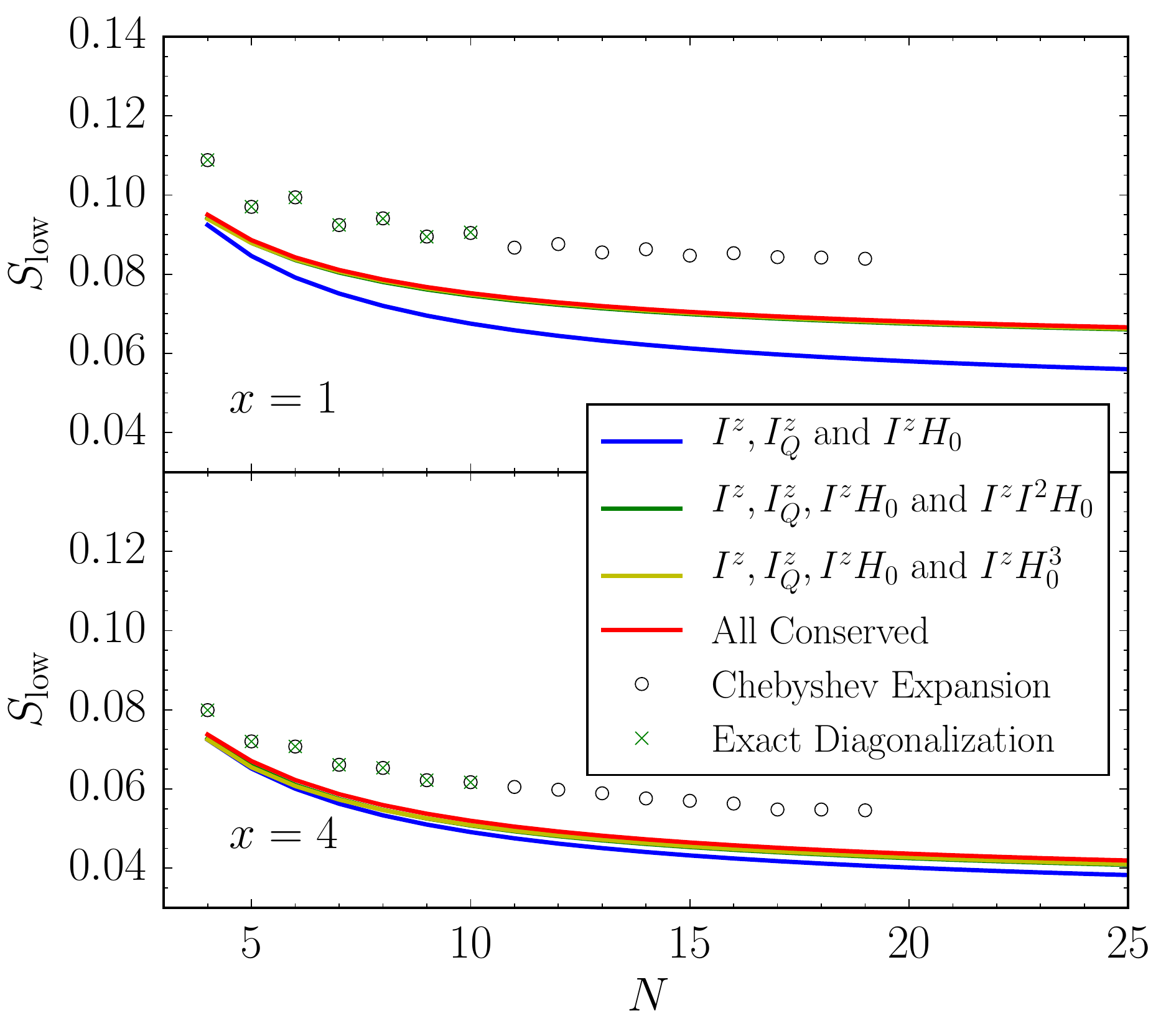}
\caption{(Color online) Improved lower bounds $S_\mathrm{low}$ using $I^z H_0^3$ (yellow) or
$I^z I^2 H_0$  (green) in addition to three constants (blue) already considered in 
Ref.\ \onlinecite{uhrig14a}.
The lower bound stemming from the combination of all these conserved quantities is depicted
in red. All curves are obtained for the coupling distribution \eqref{eq:couplings_exp}.
\label{fig:improved}}
\end{figure}

The lower bounds obtained by combining several conserved quantities are shown in 
Fig.\ \ref{fig:improved}. The dark gray (blue) curve results from the
three constants of motion $I^z$, $I^z H_0$, and 
\begin{equation}
I^z_Q:=I^z \sum_{k,l=0;k<l}^N\mathbf{S}_k\cdot\mathbf{S}_l,
\end{equation} 
which corresponds essentially to $I^z I^2$.
This set of observables was already used before\cite{uhrig14a} where it was
noticed that the resulting rigorous bounds are not very tight.
For $x=1$, we observe a significant improvement if either $I^z H_0^3$ or $I^z I^2 H_0$ is
added to the set  of conserved quantities. But including $I^z I^2$ or $I^z H_0^3$
 additionally yields only a minor improvement of the lower bound compared to 
numerical results obtained by  Chebyshev polynomial expansion  
\cite{talez84,dobro03a,dobro03b,hackm14a,pb_hackmann_2015}.

The numerical results are verified for small systems by performing an exact diagonalization
to determine all eigenenergies and eigenstates which allow us to compute 
the persisting correlation according to  Eq.\ 2 in Ref.\ \cite{uhrig14a}. 
The lower panel of  Fig.\ \ref{fig:improved} shows that the qualitative observations for $x=1$ 
also apply to $x=4$, i.e., for a much larger spread of about $\exp(-4)$. But we note that
the improvement due to the constants of motion built from higher powers of $H_0$
is not so important anymore.

Turning back to Fig.\ \ref{fig:compare_singles}, we see that the lower bounds obtained by using 
the constants of motion $I^z H_0^3$ and $I^z I^2 H_0$ alone seem to converge to the same value for
$N \to \infty$.
Indeed, determining  the leading orders in $N$ of the respective matrix and vector 
elements, we find
\begin{subequations}
\begin{eqnarray}
  \lim_{N\to\infty} S_\mathrm{low}^{(I^z H_0^3)}  =   
	\lim_{N\to\infty} \frac{1}{4}  \frac{5}{42+21 N \Sigma_2 / \Sigma_1^2}
	\\
	\lim_{N\to\infty} S_\mathrm{low}^{(I^z I^2 H_0)}  =   
	\lim_{N\to\infty} \frac{1}{4}  \frac{5}{42+21 N \Sigma_2 / \Sigma_1^2}
\end{eqnarray}
\end{subequations}
which explains the observed behavior. Hence, in the vector space of operators 
$I^z H_0^3$ and $I^z I^2 H_0$ can be assumed to be parallel for $N\to\infty$.

\begin{figure}[htb]
  \includegraphics[width=1\columnwidth,clip]{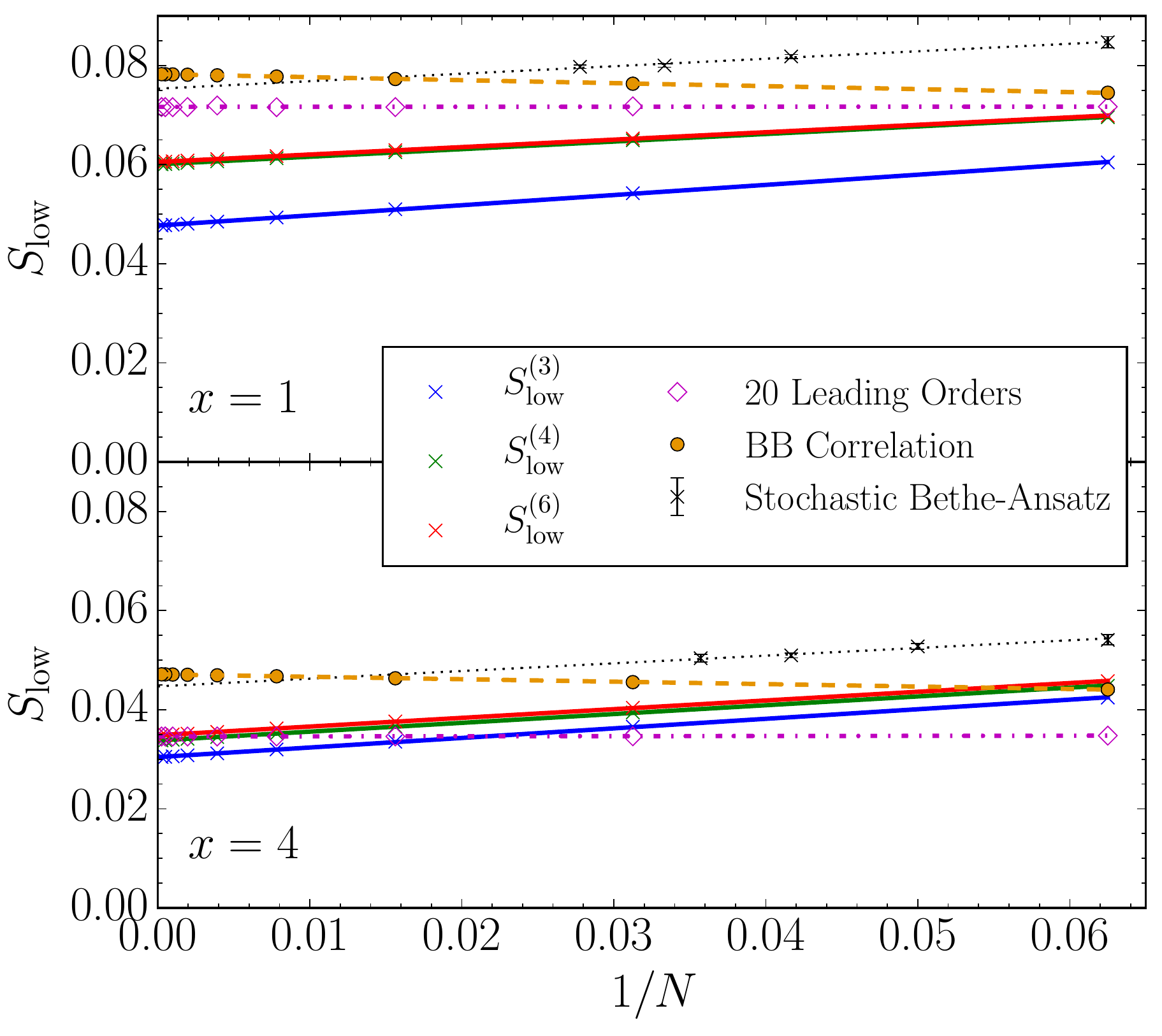}
\caption{(Color online) Extrapolated improved lower bounds $S_\mathrm{low}$. The lower bound
 obtained by using $I^z$, $I_Q^z$ and $I^z H_0$ is labeled $S_\mathrm{low}^{(3)}$. 
The bound $S_\mathrm{low}^{(4)}$ denotes the result from adding $I^z H_0^3$ and 
$S_\mathrm{low}^{(6)}$ shows the bound obtained on inclusion of $I^z H_0^2$  and $I^z I^2 H_0$. The couplings are distributed according to 
\eqref{eq:couplings_exp}. Results from stochastic Bethe ansatz are included as well
as the approximate lower bounds from the field-field correlations (BB). 
The diamonds denote the bounds obtained from the Gaussian approximation applied
to constants of motion containing high powers of $H_0$. This approximation becomes
exact upon $N\to\infty$ (see main text).
\label{fig:improved_infty}}
\end{figure}

In Fig.\ \ref{fig:improved_infty}, the lower bounds obtained by using the additional 
relevant conserved quantities are extrapolated for $1/N \to 0$.
For comparison the stochastically evaluated data from Bethe ansatz are 
included as well.
The previously observed qualitative behavior is also found in the thermodynamic limit.
For larger $x$ the rigorous lower bounds are all fairly close. They are closer to 
the Bethe ansatz data for smaller $x$. For larger $x$ it appears that some
important relevant constants of motion are still missing.
We also note that $S_\mathrm{low}^{(4)}$ and $S_\mathrm{low}^{(6)}$ appear to converge to the same value for $N \to \infty$. This can be understood by the above observation
that $I^z I^2 H_0$ and $I^z H_0^3$ are parallel in operator space for $N \to \infty$.

\subsection{Analytical results for infinite spin bath}
\label{subsec:infin_bath}

In the previous section, we saw a significant improvement of the lower bounds 
upon including the constant of motion $I^z H_0^3$. Yet, the resulting rigorous bounds were
still far from tight. Hence, it suggests itself to generalize $I^z H_0^3$
to arbitrary powers of $H_0$. The question is to which extent the inclusion of 
higher powers yields additional information about the spin dynamics, which
is linearly independent of the conserved lower powers.
Thus, we consider conserved quantities of the form $I^z H_0^{2m-1}$ with $m \in \mathbb{N}$.
But, because the evaluation of matrix elements for $I^z H_0^5$, let alone even higher powers of 
$H_0$, is limited by the exponential increase of the runtime of the computer
algorithms, we resort to studying the matrix and vector elements in the 
 leading order in $N$. This approach has two advantages. First, we can directly
address the  infinite spin bath and, second, the calculations are
decisively simpler.

In Sec.\ \ref{sec:id_relevant}, we observed that the leading order in $N$ of a 
scalar product of operators is obtained by maximizing the number of 
contractions which directly implies to compute all pairwise contractions.
This finding corresponds to the observation by \citeauthor{cywin09b} that the pairing of spin operators leads to the leading order in the diagrammatic expansion of the decoherence function 
$W(t) = \braket{S^\dagger_{-}(t) S_{+}(t)}$ \cite{cywin09b}.

Since we only have to consider the pairwise contractions of equal components of spin operators
we can do so by Gaussian integrals.  It is well known that the evaluation of expectation values
of products of variables which follow Gaussian distributions amounts to the
computation of all pairwise contractions \cite{isser18,wick50}. Thus, we may
treat each spin in the bath as classical vector of which the components are
Gaussian distributed with variance $1/4$ because we consider
$S=1/2$ and set $\hbar=1$. 
We emphasize that we do not approximate the quantum spins by classical vectors.
But we compute the leading order in $N$ of traces over the Hilbert space of $N$ quantum 
spins using Gaussian integrals.

Concretely, we consider  matrix elements of the form 
$(I^z | I^z H_0^{2m})$ with $m \in \mathbb{N}$ which cover 
also all combinations $(I^z H_0^{m-n} | I^z H_0^{m+n})=(I^z | I^z H_0^{2m})$
for $n\le m$ due to the Hermiticity of $H_0$. To evaluate these matrix elements,
it is sufficient to treat $\mathbf{B}$ and $I^z$ as Gaussian distributed variables,
but with a certain correlation between them. The details of the calculation
are given in Appendix \ref{app:analytical_gauss}; the final result reads as
\begin{eqnarray}
  \label{eq:theorem_leading1}
  &&(I^z | I^z H_0^{2m}) \ = 
	\\ \nonumber
	&& \quad \frac{(2m+1)!!}{2^{4m+2}} \left( N \Sigma_2^m + \frac{2m}{3} 
	\Sigma_2^{m-1} \Sigma_1^2 \right) + \mathcal{O}(N^m).
\end{eqnarray}
For the vector element $(S_0^z | I^z H_0^{2m-1})$ we obtain in an analogous way
\begin{equation}
  \label{eq:theorem_leading2}
  (S_0^z | I^z H_0^{2m-1}) = \frac{1}{2^{4m}} \frac{(2m+1)!!}{3} \ \Sigma_2^{m-1} \Sigma_1 + \mathcal{O}(N^{m-1}).
\end{equation}

\begin{figure}[htb]
	\includegraphics[width=1.0\columnwidth]{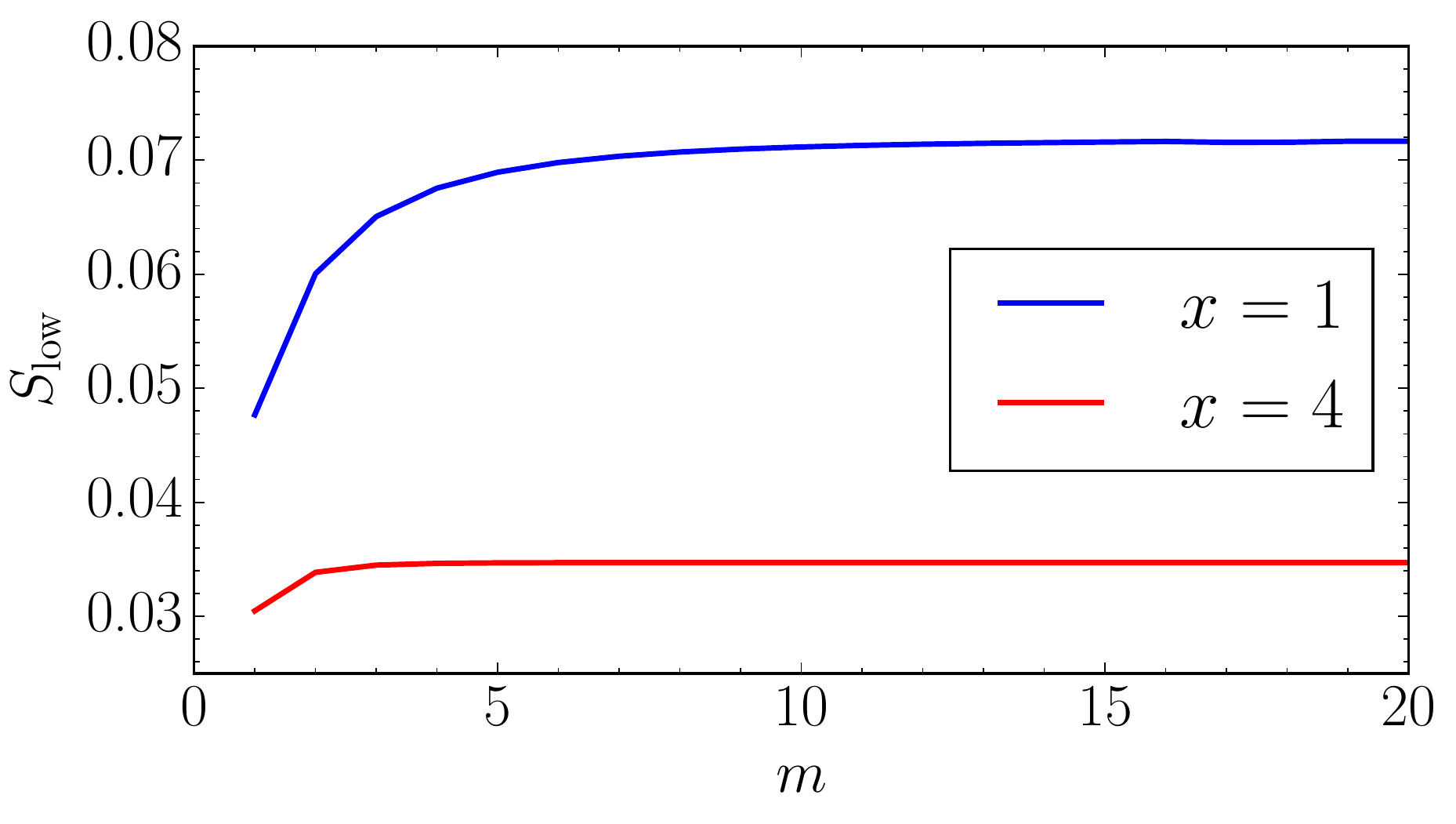}
  \caption{(Color online) Convergence of $S_\mathrm{low}^{(m)}$ as a function of $m$ for the 
	constants of motion $\{I^z H_0, \dots, I^z H_0^{2m-1}\}$ for exponential couplings 
	given by \eqref{eq:couplings_exp} and a bath size of $N=20$. Larger values of $N$ do
	not change the curves significantly. 
		\label{fig:leadingOrders}}
\end{figure}

With the general formulas for the leading orders in $N$ of matrix and vector elements at our
disposal,  we can proceed to combine an arbitrary number of constants of motion
of type $I^z H_0^{2m-1}$ and compute lower bounds according to 
\eqref{eq:slow} in leading order of $N$, i.e., for infinite spin bath.
We find that $S_\mathrm{low}$ converges quickly as a function of $m\to\infty$ to an asymptotic value as shown in Fig.\ \ref{fig:leadingOrders}.
We conclude that higher powers of a conserved quantity help to increase the lower bounds
and make them tighter. For instance, the inclusion of $I^z H_0^{3}$ makes sense.
 But the effect is not very large and seems to decrease for larger spread,
i.e., for larger values of $x$. 

Hence we consider the set of $20$ conserved quantities $\{ I^z H_0, \dots, I^z H_0^{39} \}$
using matrix and vector elements in leading orders of $N$ to compute the data points
shown by diamonds  in Fig.\ \ref{fig:improved_infty}. They can be extrapolated
reliably to $N=\infty$ where the approximation of the leading order in $N$ becomes exact.
We note a significant improvement of the lower bound using this approach,
in particular for small $x$. For small systems and larger $x$ (see lower panel of Fig.\ 
\ref{fig:improved_infty}), the leading order bound is lower than the 
bound obtained in Ref.\ \onlinecite{uhrig14a} using $I^z$, $I_Q^z$ and $I^z H_0^3$.
But, it must be kept in mind that the formulas \eqref{eq:theorem_leading1}
and \eqref{eq:theorem_leading2} are justified only for large baths.

Summarizing, we note that the inclusion of $I^z H_0^3$ or $I^z I^2 H_0$ in the set of conserved quantities yields a significantly improved lower bound. Still, the resulting bounds
do not exhaust the numerically data. Thus, further searches for the missing important
constants of motion are called for.
Furthermore, we found that $I^z H_0^3$ or $I^z I^2 H_0$ are almost parallel in the vector space of operators for large spin bath.

A technical key result is that the leading order in $N$ of scalar products can be obtained from Gaussian integrals.
This allowed us to derive a closed analytical expression for the leading orders for matrix and vector elements relevant for constants of motion of the form $I^z H_0^{2m-1}$.
Quantities of this form seem to account for a significant part of the persisting
correlations. Still, even the extrapolated bounds are not tight.

\section{Finite magnetic fields} 
\label{sec:finmag}

Next, we  study the influence of a finite external magnetic field $h$ applied to the 
central spin on the spin-spin correlation in model  \eqref{eq:h0h}.
Thus, there is no spin isotropy anymore in contrast to the situation dealt with
in the previous sections. Therefore, only the $S_0^z$-$S_0^z$ correlation
can be expected to display a persisting portion.
The correlations  for $S_0^x$ as well as $S_0^y$ vanish for finite magnetic field strength
due to Larmor precession.

First, we  provide an analytical lower bound allowing for easy and fast verification of numerically obtained results. Then, we use the model's integrability to improve lower bounds
 and we extend our results to constants of motion composed of  quadratic and cubic powers of 
$H_0(h)$. Finally, the best lower bounds are compared to numerical results for small bath sizes 
to assess how tight the bounds are.

\subsection{Rigorous spin-spin bounds} 
\label{subsec:ss_h}

On the basis of the three conserved operators $H_0(h)$, $I^z$, and $H_0^z(h):=I^z H_0(h)$ we  calculate lower bounds depending on the strength $h$ of the external magnetic field. 
Note that $H_0(h)$ contrary to $H_0$ overlaps with the operator of interest $S_0^z$ 
for $h \neq 0$. Instead of the aforementioned technique of using spin operator contractions for 
$S_i^\alpha$ in order to calculate the required scalar products, it is also possible to rewrite
 the conserved quantities in terms of ladder operators $S^\pm$.

Considering only $H_0^z(h)$ leads to a lower bound $S_\mathrm{low}\propto h^{-2}$ for $h\to\infty$ 
at given fixed bath size. This is physically unreasonable since one expects
a better and better protection of the $z$-magnetization for a larger and larger
field. Hence we choose a set of two constants of motion  $H_0^z(h)$ and $H_0(h)$. 
The resulting bound can be extrapolated for the exponentially distributed couplings 
\eqref{eq:couplings_exp} to infinite number of bath spins yielding
\begin{equation}
	\label{eq:s_infty_h}
	S_{\infty}=\frac{\splitfrac{e^x \left(8 h^4 x+h^2 (6x-4)+3\right)}{+8 h^4 x+h^2 (6x+4)-3}}{\splitfrac{2e^x \left(16 h^4 x+8 h^2 (3x-2)+9 x+12\right)}{+2 \left(16 h^4 x+8 h^2 (3x+2)+9 x-12
	\right)}}.
\end{equation}
Here we exploit the analytically accessible  expressions for $\Sigma_m$. In order to 
have well-defined limits of the $\Sigma_m$ for $N\to\infty$ we normalize 
$\Sigma_2=J_Q^2=1$ by choosing
\begin{equation}
	J:=\sqrt{\frac{\exp\left(2x /N\right)-1}{1-\exp \left(-2x\right)}}
\end{equation}
 in \eqref{eq:couplings_exp} and use it throughout this section.
This corresponds to using $J_Q$ as energy unit.
We express the magnetic field strength $h$ relative to this unit below. 
The analytical result \eqref{eq:s_infty_h} can be used for quick and rough checks of 
numerical results for variable spreads $x$. 

\begin{figure}[htb]
  \centering
  \includegraphics[width=1\columnwidth,clip]{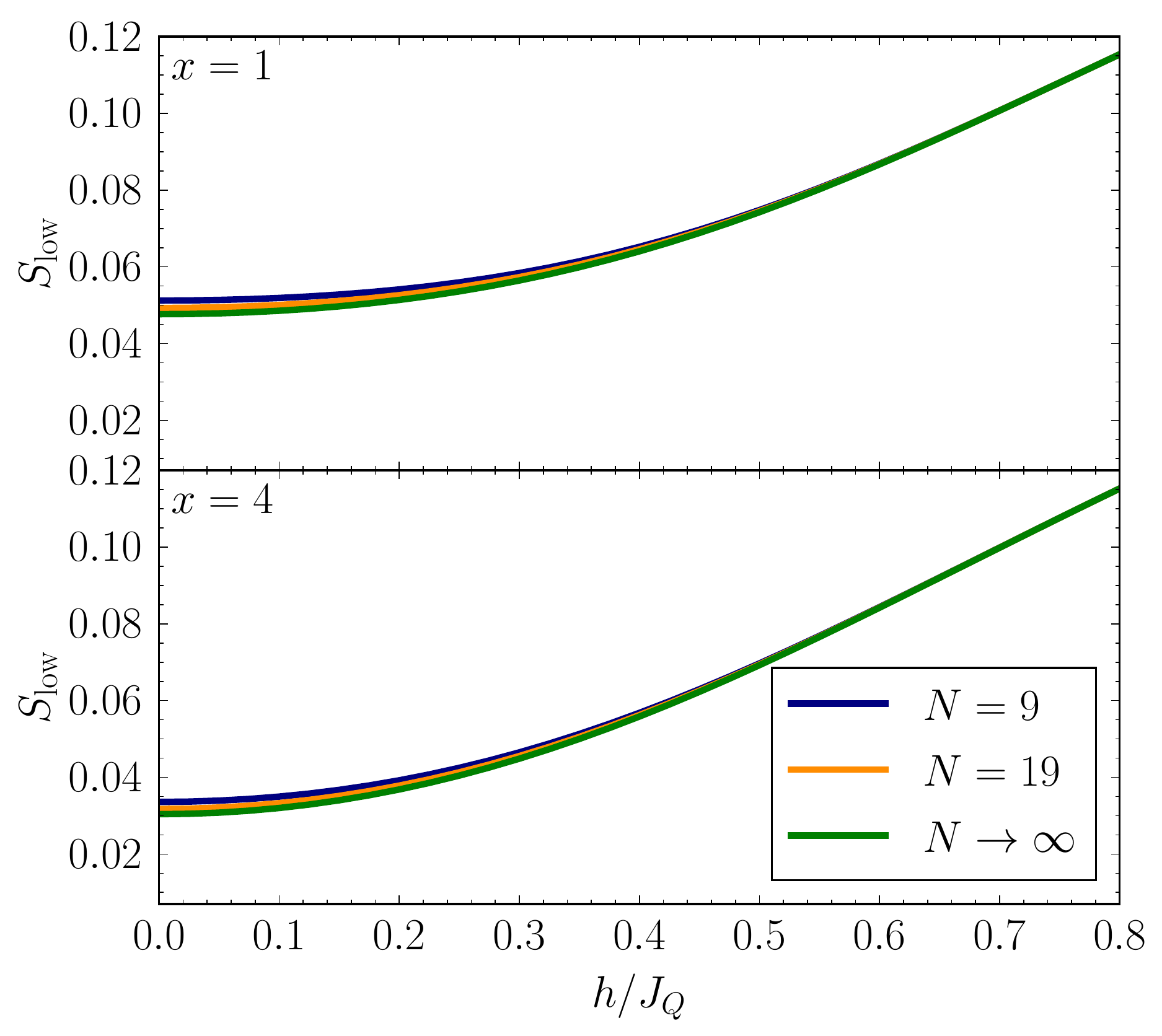}
  \caption{(Color online) Lower bounds
	and their dependence on the magnetic field $h$ for two finite bath sizes $N=9$ and 
	$N=19$ compared to the infinite bath limit for the exponential couplings in 
	\eqref{eq:couplings_exp}.}
  \label{fig:convergence_h}
\end{figure}

In Fig.\ \ref{fig:convergence_h}, we 
illustrate the very rapid convergence of the bounds for increasing $N$ for a fixed magnetic field strength $h$. The influence of the bath size decreases quickly for increasing $h$. 
The latter corresponds to the physically motivated expectation according to which the Overhauser field becomes less and less important for  rising external magnetic field.
Combining all three quantities $I^z$, $H_0^z(h)$,  and $H_0(z)$ leads to a bound that reproduces
 \eqref{eq:s_infty_h} in the infinite bath limit.

\begin{figure}[htb]
  \centering
  \includegraphics[width=1\columnwidth,clip]{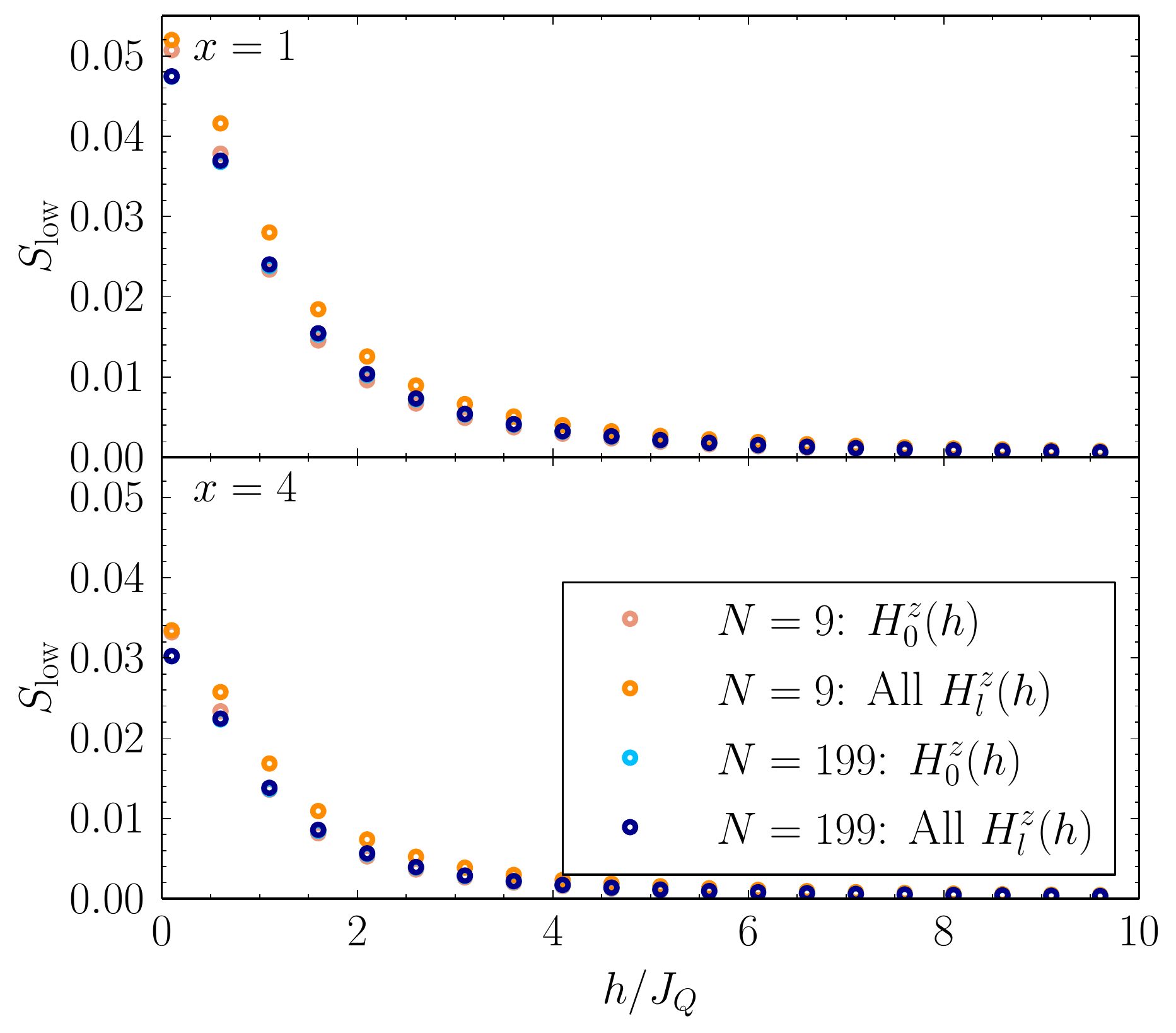}
  \caption{(Color online) Lower bounds
	and their dependence on the magnetic field $h$ for two finite bath sizes $N=9$ and 
	$N=199$ computed for the constants of motion of type $H^z_l(h)$
	for the exponential couplings in 	\eqref{eq:couplings_exp}.
	Clearly, the inclusion of all constants of motion beyond $H^z_0(h)$
	does not provide significant improvement, in particular for systems which
	are not very small.
	The bounds deteriorate for larger fields indicating that the considered constants
	of motion matter only for small fields.}
  \label{fig:compare_pb}
\end{figure}

The model's integrability \eqref{eq:hlh} provides us with $N+1$ constants of motion
$H_l(h)$ which can be used for calculating lower bounds. 
If the operators $X_i$ commute pairwise then  any product $X_i X_j$ commutes with any 
$X_k$ or $X_k X_l$. Thus, the constants of motion $H_l^z(h):=I^z H_l(h)$ commute pairwise 
and with any $H_k(h)$ leading to $2N+2$ conserved quantities at most. 
Using the scalar products listed in Appendix \ref{app:elements_list}, lower bounds for
two different bath sizes $N=9$ and $N=199$ are calculated and shown in 
Fig.\ \ref{fig:compare_pb}.
For each bath size we compare the lower bounds obtained from  $H_0^z(h)$
with the ones obtained from $N+1$ conserved quantities $H_l^z(h)$. 
Only in the range of low magnetic field strengths and small
 bath sizes, there is noticeable deviation between both results. 
Even for moderate bath sizes of $N=199$ the numerical data from one and from $N+1$
 conserved quantities agrees perfectly.

\begin{figure}[htb]
  \centering
  \includegraphics[width=1\columnwidth,clip]{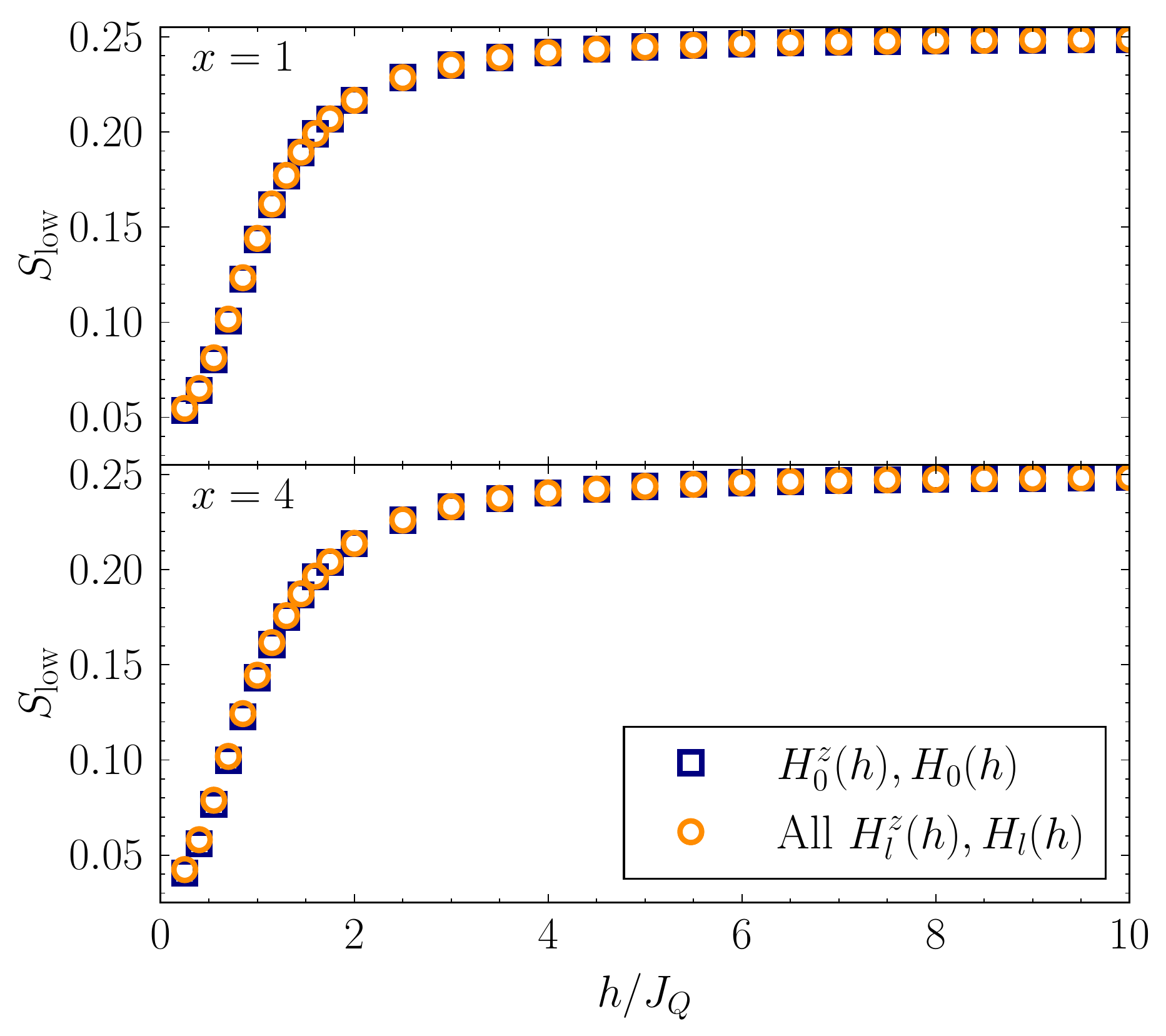}
  \caption{(Color online) Comparison of the rigorous bounds from two and $2N+2$ conserved 
	quantities for bath size $N=199$ and the exponential couplings in \eqref{eq:couplings_exp}.}
	\label{fig:integrability_h_199}
\end{figure}

We repeat this comparison  for the set of two constants of motion $H_0^z(h)$ and $H_0(h)$ 
and the set of $2N+2$ constants of motion $H_l^z(h)$ and $H_l(h)$. As depicted in 
Fig.\ \ref{fig:integrability_h_199}, exploiting the integrability has no significant impact on the corresponding lower bounds for finite magnetic fields.
This observation leads us to the conclusion that integrability  is of minor importance
in the case of  finite external magnetic fields. 
This finding extends the previous conclusion concerning the limited role
of integrability in absence of magnetic fields in Ref.\ \onlinecite{uhrig14a}.

\subsection{Improvement due to quadratic and cubic powers of $H_0(h)$}

Without magnetic field we have shown in Sect.\ \ref{sec:improved} that higher powers
of conserved quantities yield a significant improvement of the bounds. Thus,
we test this idea also in presence of magnetic fields. In particular for
small fields it is desirable to realize such improvement.
Thus, we consider the most relevant  operators $H_0^z(h)$ and $H_0(h)$ and 
extend them by quadratic and cubic terms. Taking into consideration 
the results from Sect.\ \ref{sec:computer_aided}, we assume the operator 
$I^z H_0^3(h)$ to also have a notable impact on the lower bound for finite magnetic fields.
Since the operator $H_0^3(h)$ has a non-vanishing overlap with $S_0^z$ for $h\neq0$, 
we include it likewise. To complete the analysis, we even include the operators 
$I^z H_0^2(h)$ and $H_0^2(h)$. By means of the technique described in Appendix 
\ref{subapp:algo_matrix}, we are able to 
calculate the scalar products of the operator of interest $S_0^z$ with all 
conserved quantities as well as all necessary matrix elements of the norm matrix $\mathbf{N}$.

\begin{figure}[htb]
  \centering
  \includegraphics[width=1\columnwidth,clip]{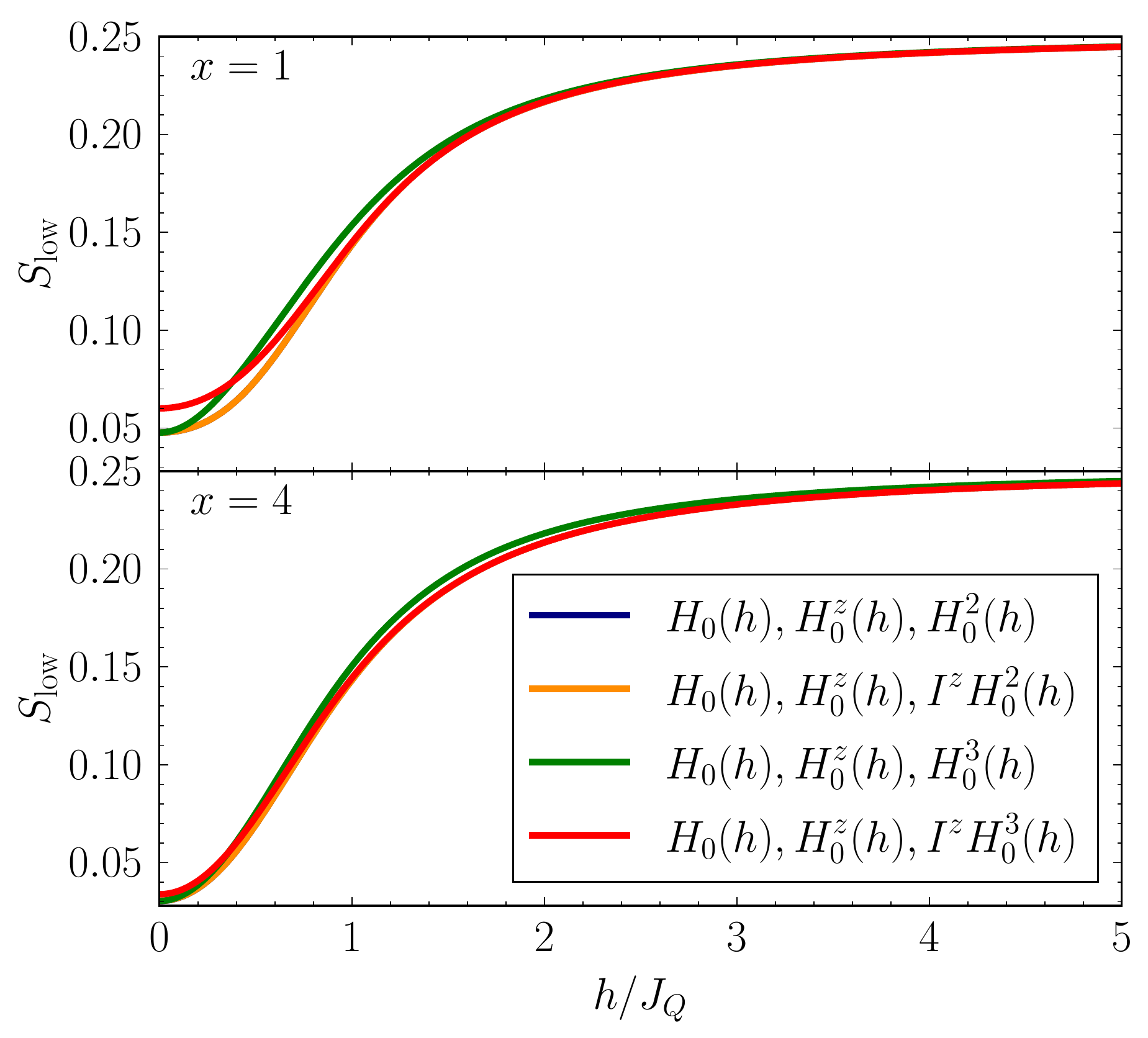}
  \caption{(Color online) Comparison of different bounds generated by combining $H_0^z(h)$ and $H_0(h)$ and a third conserved quantity as denoted in the legend for bath size $N=999$ and 
	the exponential couplings in \eqref{eq:couplings_exp}.}
	\label{fig:quadratic_cubic_h}
\end{figure}

The data in Fig.\ \ref{fig:quadratic_cubic_h} shows that the inclusion of cubic powers of 
$H_0(h)$  improves  the lower bounds significantly. We compared this result to results
 for smaller baths (not shown) and again found a quick convergence in $N$ for finite fields $h$.

\begin{figure}[htb]
  \centering
  \includegraphics[width=1\columnwidth,clip]{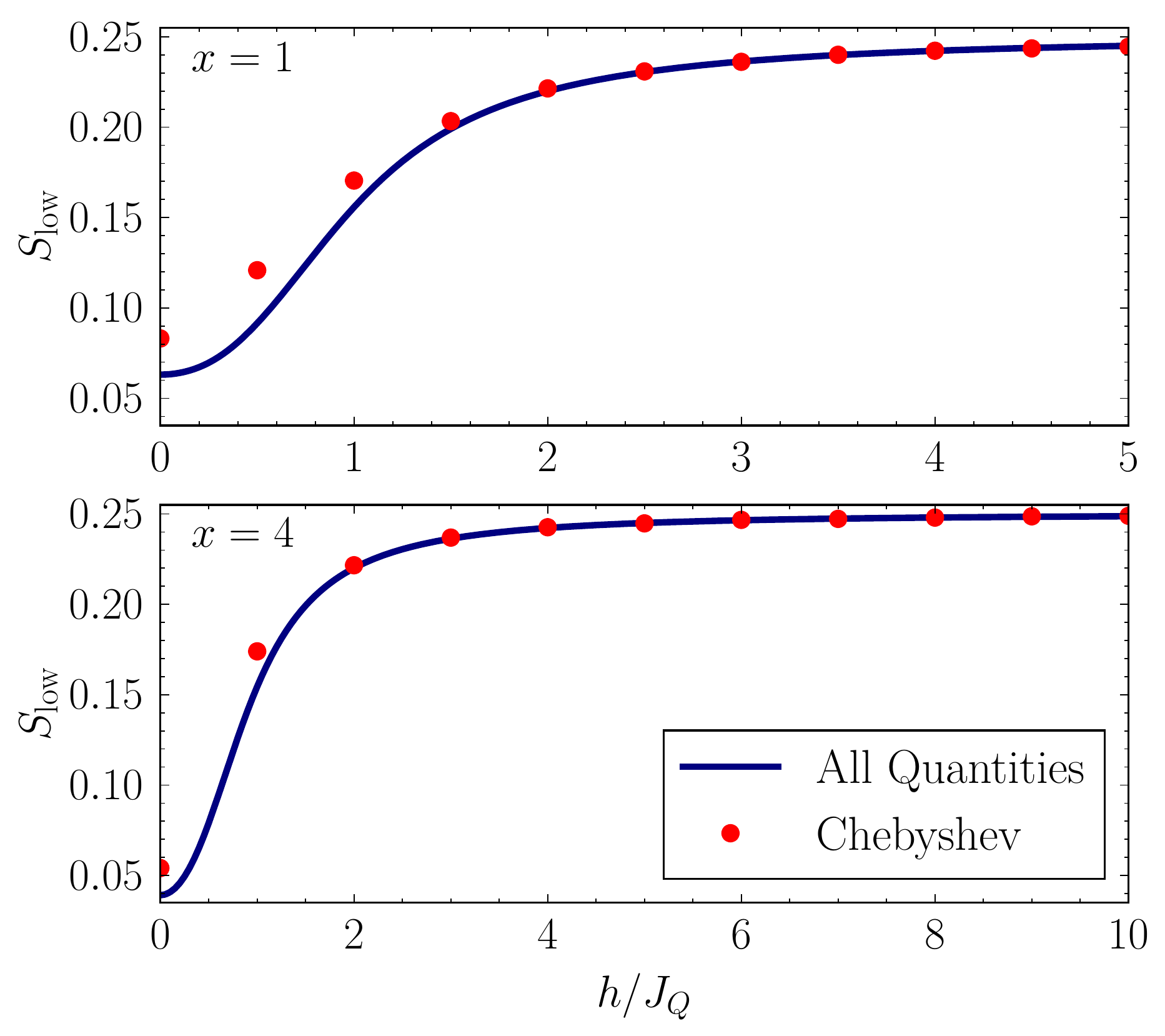}
  \caption{(Color online) Comparison of the bounds generated by combining all six conserved quantities (see legend of Fig.\ \ref{fig:quadratic_cubic_h}) to numerical data from Chebyshev polynomial expansion for bath size $N=19$ and the exponential couplings
	in \eqref{eq:couplings_exp}.}
	\label{fig:master_bound_h}
\end{figure}

To further optimize the lower bound, we combine all six conserved quantities and compare the results to numerical data \cite{pb_hackmann_2015} computed with a precision of $10^{-4}$ by
Chebyshev polynomial expansion \cite{talez84,dobro03a,dobro03b,hackm14a}. The results in 
Fig.\ \ref{fig:master_bound_h} display an excellent agreement of the bounds with the numerical
data for $h\gtrapprox 2J_Q$. Thus, our bounds are already very tight for moderate and large
fields. Note that our results clearly show that the ratio $h/J_Q$ is the relevant
one determining the qualitative behavior of the system. Previous studies often
indicated that the ratio $h/\Sigma_1$, i.e., the magnetic field over the sum
of all couplings is decisive \cite{khaet02,khaet03,coish04,coish10}.
Only for smaller fields the bounds are not very tight, although they still capture
most of the persisting correlation (note the offset on the ordinate of 
Fig.\ \ref{fig:master_bound_h}). This observation agrees with what we had seen
in the previous sections dealing with the CSM without magnetic fields.

Furthermore, finite magnetic fields suppress finite-size effects so that
moderately large spin baths are sufficient to obtain data coinciding with
data in the thermodynamic limit.

\section{Conclusions} 
\label{sec:conclusio}

Understanding coherence and decoherence in quantum many-body systems is an 
important issue to develop quantum technology.
Quantum coherent control is one of its central issues. An important model
in this field is the central spin model because it describes the decoherence of
an elementary two-level system coupled to a large environment of spins, i.e, a bath of spins.
Many numerical and approximate approaches have been introduced. But rigorous
results are still rare, in particular those referring to the long-time behavior
\cite{uhrig14a,hette15}.
Even the exact solvability by Bethe ansatz is only of limited help because
the corresponding equations are extremely difficult to treat for large
bath sizes.

In this context, this paper provides a number of rigorous results
which constitute important extensions or  improvements of previous findings
\cite{uhrig14a}. These results can serve as test bed for numerical and approximate
approaches. In particular, they allow one to make reliable statements about
the long-time behavior of extremely large spin baths which may easily comprise
$10^6$ spins or more.

We have shown that a well-defined limit $N\to\infty$
(thermodynamic limit)
exists if the moments $\Sigma_m$ have well-defined limits for $N\to\infty$
for properly scaled energy scale $\Sigma_2=J_Q^2$. 
This limit has hardly been studied so far except in Ref.\ \onlinecite{hackm14a}.
We illustrated this
for exponentially distributed couplings between  1 and $\exp(-x)$ and 
investigated the dependence  on $x$ of the lower bounds in the 
thermodynamic limit. Clearly, the persisting correlation vanishes
for $x\to\infty$. It implies that no coherence remains at all if
the central spin is coupled to spins of which the couplings
are almost all infinitesimally weak.

The rigorous lower bounds addressing directly the spin-spin correlation are not
yet tight for vanishing magnetic fields. 
So, one extension considers the approximate bounds derived from 
the field-field correlations of the Overhauser field which  appear to be tight indeed.
So, we deduced the $x$ dependence of the persisting correlation in this way and
established its  asymptotic behavior empirically. It is found to 
be given by $\ln(x)/x$. Using the heuristic replacement $x\to \ln(t)$, this
implies a refined long-time behavior with a nested logarithm not found before
\cite{khaet02,khaet03,coish04,erlin04,chen07}.

An alternative extension aims at making the the rigorous bounds 
tighter. We identified
further relevant constants of motion involving higher powers, for instance 
$I^z H_0^3$ and $I^z I^2 H_0$. Their inclusion indeed yields
a significant improvement, but no tight bounds.
To further investigate the effect of even higher powers we 
established an efficient  approach to compute the required scalar products
in the thermodynamic limit via Gaussian integrals. 
The evaluation of the resulting bounds reveals some improvement,
but still the bounds are not tight. Thus, we conclude that some important
constants of motion or products of constants of motion have still
to be identified so that further studies are called for.

Finally, we extended the rigorous approach to the experimentally relevant situation of
a finite magnetic field applied to the central spin. 
Due to the reduced symmetry we may only study the persisting magnetization
in the direction of the magnetic field. We found that already a moderately
large magnetic field of the order of the characteristic energy $J_Q$
leads to tight bounds. This confirms $J_Q$ as the relevant
internal energy scale in comparison to applied external fields. 
A small number of constants of motion suffices to
yield remarkable agreement with numerical data. For the two most relevant
constants of motion we could derive a simple analytical expression
which directly captures the $N=\infty$ limit.
Generally, the finite-size effects are
strongly suppressed as well so that moderately large baths
of about 100 spins yield bounds which almost coincide with the thermodynamic
limit. 

In conclusion, the generalized Mazur inequality \cite{mazur69,suzuk71,uhrig14a}
allows one to capture the long-time limit of the central spin model. Thereby,
the understanding of slow decoherence in this widely employed model
has been enhanced. Application to other extended models is within reach.

\begin{acknowledgments} 
We gratefully acknowledge the support of TRR 160
``Coherent manipulation of interacting spin excitations in tailored semiconductors''
of the Deutsche Forschungsgemeinschaft. We are thankful to F. Anders, A. Greilich, J. Hackmann, and J. Stolze for helpful discussions and provision of numerical data.
We also like to thank K. Dungs for providing the \texttt{cpp-paulimagic} package.
\end{acknowledgments}

\begin{appendix}

\section{Analytical calculation of Gaussian correlations}
\label{app:analytical_gauss}

In the main text in Sect.\ \ref{subsec:infin_bath} we showed that $\mathbf{B}$ and $I^z$ 
can be seen as classical Gaussian variables for
the calculation of the leading order in the bath size $N$. This does
not apply to the central spin. So we need an additional consideration
first. This is provided by Eq.\ \eqref{eq:Bsquared} which shows that 
$H_0^2$ corresponds to $B^2/4$ except for a correction which involves the
commutators of the components of $\mathbf{B}$. In Ref.\ \onlinecite{stane14b}
it was shown that this commutator contributes in lower powers of $N$ than the
$B^2$ term. Thus, we can safely replace $H_0^{2m}$ by $B^{2m}/4^m$ in leading
order in $N$ so that 
\begin{equation}
(I^z | I^z H_0^{2m}) = \braket{(I^z)^2 B^{2m}} / 2^{2m} + 
\mathcal{O}(N^m).
\end{equation}
The right hand side of this expression is evaluated by
integration over the appropriate multivariate Gaussian distribution
\begin{equation}
  \braket{f(\mathbf{x})} = \frac{ \int_{\mathbb{R}^k} f(\mathbf{x}) \exp \left[ -\frac{1}{2} 
	\mathbf{x}^T \mathbf{\Omega}^{-1} \mathbf{x} \right] \mathrm{d}^k \mathbf{x}}{\sqrt{(2 \pi)^k \mathrm{det}(\mathbf{\Omega})}},
\end{equation}
where the vector $\mathbf{x}$ is given by $\mathbf{x} = (B^x, B^y, B^z, I^z)^T$. The covariance matrix $\mathbf{\Omega}$ reads as
\begin{equation}
  \mathbf{\Omega} = \begin{pmatrix}
    \sigma^2 & 0 & 0 & 0 \\
    0 & \sigma^2 & 0 & 0 \\
    0 & 0 & \sigma^2 & \beta^2 \\
    0 & 0 & \beta^2 & \alpha^2
  \end{pmatrix},
\end{equation}
with the entries 
\begin{subequations}
\label{eq:covar_entries}
\begin{align}
\sigma^2 &= (B^z | B^z) = \Sigma_2/4
\\
\beta^2 &= (B^z | I^z) = \Sigma_1/4
\\
\alpha^2 &= (I^z | I^z) = N/4.
\end{align}
\end{subequations}
Integration over $I^z$  yields
\begin{align} 
\label{eq:integration_Iz}
  \braket{(I^z)^2 B^{2m}} &= \int_{\mathbb{R}^3}  \frac{(\alpha^2 \sigma^4 - \sigma^2 \beta^4 + \beta^4 (B^z)^2)}{\sigma^7 \sqrt{(2 \pi)^3}} B^{2m} \cdot
	\nonumber \\
  &\times \exp \left[- \frac{(B^x)^2+ (B^y)^2 + (B^z)^2}{2 \sigma^2} \right] \mathrm{d}^3 \mathbf{B}.
\end{align}
For the integration over $\mathbf{B}$ we use spherical coordinates $(B,\varphi,\theta)$.
The first two terms in \eqref{eq:integration_Iz} are easy to treat because
they are rotationally symmetric. We use
\begin{align} 
\label{eq:integral_wick1}
  \int_\mathbb{R^3} B^{2m+2} \exp \left[ - \frac{B^2}{2 \sigma^2} \right] \sin \theta \ \mathrm{d}B \ \mathrm{d}\theta \ \mathrm{d}\varphi \nonumber \\ = \sqrt{(2 \pi)^3} (2m + 1)!! \sigma^{2m+3}.
\end{align}
The third term is integrated with the help of
\begin{align} 
\label{eq:integral_wick2}
  \int_\mathbb{R^3} B^{2m+4} \exp \left[ - \frac{B^2}{2 \sigma^2} \right] \cos^2 \theta 
	\sin \theta \ \mathrm{d}B \ \mathrm{d}\theta \ \mathrm{d}\varphi 
	\nonumber \\ 
	= \frac{1}{3} \sqrt{(2 \pi)^3} (2m + 3)!! \sigma^{2m+5}.
\end{align}
Note that \eqref{eq:integral_wick1} and \eqref{eq:integral_wick2} can be derived 
by induction in $m$. The expressions are a direct consequence of Wick's/Isserlis' theorem.
The double factorial $(2k-1)!!$ counts the number of possibilities to split a set of $2k$ elements
 into pairs, i.e., into two-point contractions \cite{isser18, wick50}.

Combining these expressions yields
\begin{align}
  \braket{(I^z)^2 B^{2m}} &= (2m+1)!! \sigma^{2m} \alpha^2 \nonumber \\
 &+ \frac{2m}{3} (2m+1)!! \sigma^{2m-2} \beta^4.
\end{align}
The wanted Eq.\ \eqref{eq:theorem_leading1} results from this expression by replacing
the entries of the covariance matrix according to \eqref{eq:covar_entries}.

The analytical expression \eqref{eq:theorem_leading2} for the leading order of 
the corresponding vector element 
$(S_0^z | I^z H_0^{2m-1})$ can be obtained in an analogous way.

\section{Algorithms for the computer-aided
 evaluation of scalar products} 
\label{app:algos}

The method described in Sec.\ \ref{sec:computer_aided} requires the evaluation of a
scalar product for a given set of couplings $\mathcal{J}$. 
Two possible approaches are outlined here.

\subsection{Spin algebra} 
\label{subapp:algo_spin}

One can implement the group structure of the Pauli matrices in an object-oriented programming
language. Tensor products can  be realized through a sequence container (such as 
$\texttt{std::vector}$ in $\texttt{C++}$), with the index $i$ of the container 
representing the $i$-th spin of the bath.
The product of two tensor products can be  simplified according to 
$\sigma^\alpha \sigma^\beta = \delta_{\alpha \beta} \mathds{1} + 
i \epsilon_{\alpha \beta \gamma} \sigma^\gamma$.
For the calculation of the trace of the final tensor product, the algorithm immediately 
discards the result because it vanishes if one of the operators in the sequence container 
is not equal to the identity matrix. Furthermore, we only need to consider factors of 
$(-1)$ and $i$ in the tensor product-class because we can sum the weighted  results 
with possible prefactors after calculating the trace due to the linearity of the trace.

The group structure can either be implemented by encoding the Pauli matrices using 
primitive types (such as a $\texttt{char}$-type) and imposing certain simplification 
rules or by using templates to create the group structure which significantly increases computational efficiency \cite{dungs15}.

The algorithm has been tested for specific scalar products of the form $(I^z | I^z H_0^{2m})$
with $m$ taking values up to 5 for $N = 4$ on an eight-core machine at 
$3.70 \ \mathrm{GHz}$ with a runtime of approximately $25$ hours.

\subsection{Hermitian matrices} 
\label{subapp:algo_matrix}

Each quantum mechanical operator can be written as a Hermitian matrix and each spin operator 
$\mathbf{S}_k$ is a matrix-triple for the components $S_k^\alpha$ as defined in 
\eqref{eq:spinoperator}. For one central spin and $N$ surrounding bath spins 
the Hilbert space has dimension $d:=2^{N+1}$. Thus, each $S_k^\alpha$ is a square matrix of 
dimension $d$. All operators used in Sect.\ \ref{sec:finmag} are sums and products of the elementary spin operators $\mathbf{S}_k$ so that we can easily generate the needed conserved
 quantities by combining the appropriate sums of products of the $\mathbf{S}_k$ weighted
by the respective prefactors such as the couplings  $J_k$. 

 For instance, the Hamiltonian \eqref{eq:h0h} is a $d\times d$ square matrix 
 and can be generated by evaluating $3N$ matrix 
multiplications of $d\times d$ matrices. Then one performs $3N+1$ matrix sums of 
$d\times d$ matrices. Here the  factors $J_k$ are scalars to be multiplied with 
the  matrices resulting from the products $\mathbf{S}_0 \cdot \mathbf{S}_k$. 
The technical implementation requires extensive caching of the elementary spin operators
$S_k^\alpha$ and of the coupling constants $J_k$ in order to achieve fast computational 
processing and to avoid increased computation time by repeated calculations.

This algorithmic approach was used to evaluate scalar products up to 
$(I^z | I^z H_0^6(h))$ for $N=6$ and various sets of couplings $\mathcal{J}$ as well as 
magnetic field strengths $h$ on a four-core machine at $3.70~\mathrm{GHz}$. 
The runtime to calculate the most complex scalar product was about $20$ seconds.

\newcommand{\scalarpr}[2]{\left(#1|#2\right)}
\newcommand{\shnj}[2]{J_#2^{(#1)}} 
\newcommand{\shns}[1]{S^{(#1)}} 
\newcommand{\shnq}[1]{Q^{(#1)}} 
\newcommand{\sigm}[1]{\mathrm{\Sigma}_{#1}} 

\section{Various vector and matrix elements} 
\label{app:elements_list}

\begingroup 
\allowdisplaybreaks
The scalar products used throughout this paper are listed here. For clarity and 
in order to follow the line of arguments presented in Sects.\ \ref{sec:improved} and 
\ref{sec:finmag} we split the scalar products into those with vanishing and those with arbitrary magnetic field strength $h$. We also draw the reader's attention to the generalizations 
of the coupling constants $J_k$ and of the $\Sigma_m$ in \eqref{eq:sigma} to
\begin{subequations}
\begin{alignat}{3}
  J_j^{(l)}&:=\frac{1}{\epsilon_l-\epsilon_j} \\
  S^{(l)}&:=\sum_{\substack{j=0\\j\neq l}}^N J_j^{(l)} \\
  Q^{(l)}&:=\sum_{\substack{j=0\\j\neq l}}^N \left(J_j^{(l)}\right)^2,
\end{alignat}
\end{subequations}
$\epsilon_0=0$, $\epsilon_l=-J_l^{-1}$ and the identities 
$J_j=J_j^{(0)}$, $\sigm{1}=S^{(0)}$, and $\sigm{2}=Q^{(0)}$.

\subsection{Vector elements for $h=0$}
\subsubsection{Overlap elements with $S_0^z$}

\begin{subequations}\begin{align}
  \left(S_0^z | I^z I^2 H_0  \right) &= \frac{1}{64} \left( (5 N + 3) \Sigma_1 \right) 
	\\
  \left( S_0^z | I^z H_0^3 \right) &= \frac{1}{256} \left(5 \Sigma_1 \Sigma_2 -4 \Sigma_3 \right) 
	\\
  \left( S_0^z | I^z H_0^2 \right) &= \frac{\Sigma_2}{64}  
\end{align}
\end{subequations}

\subsubsection{Overlap elements with $B^z$}

\begin{subequations}\begin{align}
  \left( B^z | I^z I^2 H_0 \right) &= \frac{1}{64} \left( (5 N - 7) \Sigma_2 +10 \Sigma_1^2 \right) \\
  \left( B^z | I^z H_0^3 \right) &= \frac{1}{256} \left(3 \Sigma_2^2 - 8 \Sigma_1 \Sigma_3 + 6 \Sigma_4 \right) \\
  \left( B^z | I^z H_0^2 \right) &= \frac{1}{64} \left(5 \Sigma_1 \Sigma_2 - 4 \Sigma_3 \right)
\end{align}
\end{subequations}

\subsection{Vector elements for arbitrary $h$}

\begin{subequations}
\begin{alignat}{4}
  \scalarpr{S_0^z}{H_0^z(h)} &= \frac{1}{16}\sigm{1} \\
  \scalarpr{S_0^z}{H_l^z(h)} &= -\frac{1}{16}J_l,~l > 0 \\
  \scalarpr{S_0^z}{H_0(h)} &= -\frac{h}{4} \\
  \scalarpr{S_0^z}{H_l(h)} &= 0,~l > 0. \\
  \scalarpr{S_0^z}{I^z} &= \frac{1}{4} \\
  \scalarpr{S_0^z}{H_0^2(h)} &=  0 \\
  \scalarpr{S_0^z}{I^z H_0^2(h)} &=  \frac{1}{64} \sigm{2}+\frac{h^2}{16} \\
  \scalarpr{S_0^z}{H_0^3(h)} &=  -\frac{5}{64} h \sigm{2}-\frac{h^3}{16} \\
  \scalarpr{S_0^z}{I^z H_0^3(h)} &=  \frac{3}{64} h^2 \sigm{1}+\frac{5}{256} \sigm{2} \sigm{1}-\frac{1}{64} \sigm{3}
\end{alignat}
\end{subequations}

\begin{widetext}

\subsection{Matrix elements for $h=0$}

{\allowdisplaybreaks[1]

\subsubsection{For $I^z I^2 H_0$}

\begin{subequations}
\begin{align}
  \left(I^z I^2 H_0 | I^z I^2 H_0 \right) &= \frac{1}{1024} \left( (105 N^3 -465 N^2 +687 N -327) \Sigma_2 + (210 N^2 -200 N+118) \Sigma_1^2 \right) \\
  \left(I^z I^2 H_0 | I^z H_0^3 \right) = \begin{split}\frac{1}{2^{12}} \left(40 \Sigma_1^4 + (200 N -440) \Sigma_1^2 \Sigma_2  + (-320 N + 704) \Sigma_1  \Sigma_3 \right. \\ \left. + (75N^2 -320 N +413) \Sigma_2^2 + (30 N^2 +208 N -574) \Sigma_4 \right) \end{split} \\  
  \left(I^z I^2 H_0 | I^z H_0^2 \right) &= \frac{1}{1024} \left(40 \Sigma_1^3 + (60 N -132) \Sigma_1 \Sigma_2 + (-30 N^2 +32 N +46) \Sigma_3\right) \\
  \left(I^z I^2 H_0 | I^z H_0 \right) &= \frac{1}{256} \left(\left(20 N -4 \right) \Sigma_1^2 + \left(15 N^2 -36 N +21 \right) \Sigma_2 \right) \\
  \left(I^z I^2 H_0 | I^z_Q \right ) &= \frac{1}{256} \left((75 N^2 -124 N +65 ) \Sigma_1\right) \\
  \left(I^z I^2 H_0 | I^z \right) &= \frac{1}{64} \left((20 N -4) \Sigma_1 \right)
\end{align}
\end{subequations}

\subsubsection{For $I^z H_0^3$}

\begin{subequations}
\begin{align}
  \left( I^z H_0^3 | I^z H_0^3 \right) = \begin{split} \frac{1}{2^{14}} \left(210 \Sigma_1^2 \Sigma_2^2 + \left(105 N - 317 \right) \Sigma_2^3 + \left(-18 N + 642 \right) \Sigma_4 \Sigma_2 -576 \Sigma_1 \Sigma_2 \Sigma_3 \right. \\ \left. -12 \Sigma_1^2 \Sigma_4   + \left(48 N + 128 \right) \Sigma_3^2  +192 \Sigma_1 \Sigma_5 +\left(48 N - 448\right) \Sigma_6 \right) \end{split} \\
  \left( I^z H_0^2 | I^z H_0^3 \right) &= \frac{1}{2^{12}} \left(-60 N \Sigma_2 \Sigma_3 +30 \Sigma_1 \Sigma_2^2 -40 \Sigma_1^2 \Sigma_3 +60 \Sigma_1 \Sigma_4 + 60 \Sigma_2 \Sigma_3 -48 \Sigma_5 \right) \\
  \left( I^z H_0^3 | I^z H_0 \right) &= \left( I^z H_0^2 | I^z H_0^2 \right) \\
  \left( I^z H_0^3 | I^z_Q \right) &= \frac{1}{1024} \left( \left(-6 N^2 + 22 N +20 \right) \Sigma_3 + \left(21 N - 75 \right) \Sigma_1 \Sigma_2 + 20 \Sigma_1^3 \right) \\
  \left( I^z H_0^3 | I^z \right) &= \left(I^z H_0^2 | I^z H_0 \right)
\end{align}
\end{subequations}

\subsubsection{For $I^z H_0^2$}

\begin{subequations}
\begin{align}
  \left( I^z H_0^2 | I^z H_0^2 \right) &= \frac{1}{1024} \left(20 \Sigma_2 \Sigma_1^2 + (15N-29) \Sigma_2^2 -32 \Sigma_3 \Sigma_1   + (6 N + 22) \Sigma_4 \right)  \\
  \left(I^z H_0^2 | I^z H_0 \right) &= \frac{1}{256} \left((- 6N+2) \Sigma_3 + 6 \Sigma_1 \Sigma_2\right) \\
  \left(I^z H_0^2 | I_Q^z \right) &= \frac{1}{256} \left((7N-5) \Sigma_1^2 + 3(N(N-6)+5) \Sigma_2 \right) \\
  \left(I^z H_0^2 | I^z \right) &= \left(I^z H_0 | I^z H_0 \right)
\end{align}
\end{subequations}}

\subsection{Matrix elements for arbitrary $h$}
\subsubsection{Integrability exploitation and first order quantities}

\begin{subequations}
\begin{alignat}{4}
  \scalarpr{H_l^z(h)}{H_l^z(h)} &= \frac{2}{64}\left(\shns{l}\right)^2+\frac{3}{64}(N-1)\shnq{l}+\frac{h^2}{16}(N+1) \\
  \scalarpr{H_l(h)}{H_l(h)} &=\frac{3}{16}\shnq{l}+\frac{h^2}{4} \\
  \scalarpr{H_l^z(h)}{H_p^z(h)} &= \frac{1}{16}\shnj{l}{p}\left(\shns{p}-\shns{l}\right)-\frac{3}{64}\left(N-3\right)\left(\shnj{l}{p}\right)^2+\frac{h^2}{8},~l\neq p\\
  \scalarpr{H_l(h)}{H_p(h)} &=-\frac{3}{16}\left(\shnj{l}{p}\right)^2,~l\neq p\\
  \scalarpr{H_l(h)}{H_l^z(h)} &= -\frac{h}{8}\shns{l} \\
  \scalarpr{H_l(h)}{H_p^z(h)} &= 0,~l\neq p \\
  \scalarpr{I^z}{I^z} &= \frac{1}{4}(N+1) \\
  \scalarpr{H_0^z(h)}{I^z} &= \frac{1}{8}\sigm{1} \\
  \scalarpr{H_0(h)}{I^z} &= -\frac{h}{4}
\end{alignat}
\end{subequations}

\subsubsection{For $H_0^2(h)$}
\begin{subequations}
\begin{alignat}{4}
  \scalarpr{I^z H_0(h)}{H_0^2(h)} &= -\frac{3}{64} h \sigm{2}-\frac{h^3}{16} \\
  \scalarpr{H_0(h)}{H_0^2(h)} &=  -\frac{3}{32} \sigm{3} \\
  \scalarpr{H_0^2(h)}{H_0^2(h)} &= \frac{5}{32} h^2 \sigm{2}+\frac{15}{256} \sigm{2}^2+\frac{3}{128} \sigm{4}+\frac{h^4}{16}
\end{alignat}
\end{subequations}

\subsubsection{For $I^z H_0^2(h)$}
\begin{subequations}
\begin{alignat}{4}
  \scalarpr{I^z H_0(h)}{I^z H_0^2(h)} &=  \frac{3}{32} \sigm{1} h^2+\frac{3}{128} \sigm{1} \sigm{2}+\frac{1}{128} \sigm{3}-\frac{3}{128} N \sigm{3} \\
  \scalarpr{H_0(h)}{I^z H_0^2(h)} &=  \scalarpr{I^z H_0(h)}{H_0^2(h)}\\
  \scalarpr{I^z H_0^2(h)}{I^z H_0^2(h)} &= \frac{3}{64} h^2 \sigm{1}^2+\frac{5}{128}h^2\sigm{2}(N-1)+\frac{5}{256} \sigm{1}^2 \sigm{2}+\frac{1}{1024}(15N-29)\sigm{2}^2\notag\\&-\frac{1}{32} \sigm{1} \sigm{3}+\frac{1}{512}\left(11+3 N\right) \sigm{4}+\frac{1}{64} h^4 (N+1) \\
  \scalarpr{H_0^2(h)}{I^z H_0^2(h)} &= \scalarpr{H_0(h)}{I^z H_0^3(h)}
\end{alignat}
\end{subequations}

\subsubsection{For $H_0^3(h)$}

\begin{subequations}
\begin{alignat}{4}
  \scalarpr{I^z H_0(h)}{H_0^3(h)} &=  -\frac{1}{16} h^3 \sigm{1}-\frac{5}{64} h \sigm{2} \sigm{1}+\frac{1}{16} h \sigm{3}\\
  \scalarpr{H_0(h)}{H_0^3(h)} &= \scalarpr{H_0^2(h)}{H_0^2(h)}\\
  \scalarpr{H_0^3(h)}{H_0^3(h)} &= \frac{21}{256} h^4 \sigm{2}+\frac{105 h^2 \sigm{2}^2}{1024}-\frac{3}{512} h^2 \sigm{4}+\frac{105 \sigm{2}^3}{4096}\notag\\&+\frac{3}{256} \sigm{3}^2+\frac{3}{256} \sigm{6}-\frac{9 \sigm{4} \sigm{2}}{2048}+\frac{h^6}{64} \\
  \scalarpr{H_0^2(h)}{H_0^3(h)} &= -\frac{5}{64}h^2\sigm{3}-\frac{15}{256}\sigm{2}\sigm{3} \\
  \scalarpr{I^z H_0^2(h)}{H_0^3(h)} &= \scalarpr{H_0^2(h)}{I^z H_0^3(h)}
\end{alignat}
\end{subequations}

\subsubsection{For $I^z H_0^3(h)$}
\begin{subequations}
\begin{alignat}{4}
  \scalarpr{I^z H_0(h)}{I^z H_0^3(h)} &= \scalarpr{I^z H_0^2(h)}{I^z H_0^2(h)}\\
  \scalarpr{H_0(h)}{I^z H_0^3(h)} &=  \scalarpr{I^z H_0(h)}{H_0^3(h)} \\
  \scalarpr{I^z H_0^3(h)}{I^z H_0^3(h)} &= \frac{15}{512} h^4 \sigm{1}^2+\frac{21(N-1)}{1024}h^4\sigm{2}-\frac{3h^2 \sigm{3} \sigm{1}}{32}+\frac{63 h^2 \sigm{2} \sigm{1}^2}{1024}\notag\\&+\frac{105N-267}{4096}h^2\sigm{2}^2+\frac{165-3N}{2048}h^2\sigm{4}+\frac{321 \sigm{4} \sigm{2}}{8192}+\frac{105 \sigm{1}^2 \sigm{2}^2}{8192}\notag\\&-\frac{9\sigm{1} \sigm{2} \sigm{3}}{256}+\frac{\sigm{3}^2}{128}+\frac{3 N \sigm{3}^2}{1024}+\frac{3}{256} \sigm{1} \sigm{5}-\frac{7\sigm{6}}{256}+\frac{3 N \sigm{6}}{1024}\notag\\&-\frac{3 \sigm{1}^2 \sigm{4}}{4096}-\frac{9 N \sigm{4} \sigm{2}}{8192}+\frac{(105N-317)\sigm{2}^3}{16386}+\frac{h^6 (N+1)}{256} \\
  \scalarpr{H_0^2(h)}{I^z H_0^3(h)} &=  -\frac{5}{128} h^3 \sigm{2}+\frac{5}{128} h \sigm{1} \sigm{3}-\frac{15}{512} h \sigm{4}-\frac{15 h \sigm{2}^2}{1024}-\frac{h^5}{64} \\
  \scalarpr{I^z H_0^2(h)}{I^z H_0^3(h)} &= \frac{5}{128} h^4 \sigm{1}+\frac{15}{256} h^2 \sigm{1} \sigm{2}-\frac{5}{256} (N+1)h^2 \sigm{3}+\frac{15 \sigm{1} \sigm{2}^2}{2048}\notag\\&-\frac{5}{512} \sigm{1}^2 \sigm{3}+\frac{15 \sigm{1} \sigm{4}}{1024}-\frac{3}{256} \sigm{5}-\frac{15}{1024} (N-1) \sigm{2} \sigm{3} \\
  \scalarpr{H_0^3(h)}{I^z H_0^3(h)} &= -\frac{3}{128} h^5 \sigm{1}-\frac{21}{256} h^3 \sigm{1} \sigm{2}+\frac{1}{16} h^3 \sigm{3}+\frac{9}{128} h \sigm{2} \sigm{3}\notag\\&+\frac{3 h \sigm{1} \sigm{4}}{1024}-\frac{3}{128} h \sigm{5}-\frac{105 h \sigm{1} \sigm{2}^2}{2048}
\end{alignat}
\end{subequations}
\end{widetext}
\endgroup
\end{appendix}


\end{document}